\documentclass[12pt]{amsart}
\usepackage{latexsym}
\usepackage{amssymb}
\usepackage{amsxtra}
\usepackage{amsmath}
\usepackage{amsfonts}

\newcommand {\sectionnew}[1]{\section{#1}}
\numberwithin{equation}{section}
\newcommand{\beq}{\begin{equation}}
\newcommand{\eeq}{\end{equation}}
\newcommand{\beqa}{\begin{eqnarray}}
\newcommand{\eeqa}{\end{eqnarray}}
\newcommand{\beaa}{\begin{eqnarray*}}
\newcommand{\ben}{\begin{eqnarray*}}
\newcommand{\eaa}{\end{eqnarray*}}
\newcommand{\een}{\end{eqnarray*}}

\newcommand \nc {\newcommand}

\newtheorem{theorem}{Theorem}[section]
\newtheorem{lemma}[theorem]{Lemma}

\newtheorem{corollary}[theorem]{Corollary}

\nc \thref{Theorem \ref}
\nc \leref{Lemma \ref}
\nc \prref{Proposition \ref}
\nc \coref{Corollary \ref}
\nc \deref{Definition \ref}
\nc \exref{Example \ref}
\nc \reref{Remark \ref}

\newcommand{\leftexp}[2]{{\vphantom{#2}}^{#1}{#2}}

\newcommand{\W}{\mathcal{W}}
\newcommand{\A}{\mathcal{A}}

\newcommand{\C}{\mathbb{C}}
\newcommand{\D}{\mathcal{D}}

\newcommand{\F}{\mathcal{F}}
\renewcommand{\H}{\mathcal{H}}

\renewcommand{\L}{\mathcal{L}}
\newcommand{\M}{\mathcal{M}}

\renewcommand{\O}{\mathcal{O}}
\newcommand{\QQ}{\mathbb{Q}}
\newcommand{\R}{\mathbb{R}}
\newcommand{\T}{\mathcal{T}}

\newcommand{\Z}{\mathbb{Z}}

\newcommand{\f}{\mathbf{f}}

\renewcommand{\k}{\mathfrak{k}}
\newcommand{\p}{\mathbf{p}}
\newcommand{\q}{\mathbf{q}}
\renewcommand{\t}{\mathbf{t}}

\def\d{\partial}

\def\iso{\cong}
\def\tensor{\otimes}

\def\Poincare{Poincar\'e}

\def\({\left(}
\def\){\right)}
\def\[{\left[}
\def\]{\right]}
\def\<{\left\langle}
\def\>{\right\rangle}

\def\gl{\lambda}
\def\ge{\epsilon}
\def\ga{\alpha}

\def\gb{\beta}
\begin{document}
\title{$\mathcal{W}_{N+1}$-constraints for Singularities of Type $A_N$}
\author{Bojko Bakalov}
\address{Department of Mathematics\\
North Carolina State University\\
Raleigh, NC 27695, USA\\
e-mail: bojko\_bakalov@ncsu.edu}

\author{Todor Milanov}
\address{Department of Mathematics\\
North Carolina State University\\
Raleigh, NC 27695, USA\\
e-mail: temilano@ncsu.edu}

\thanks{The first author is supported in part by the NSF grant DMS-0701011} 
\thanks{
The second author is supported in part by the NSF grant DMS-0707150.}
\date{\today}

\begin{abstract}
Using Picard--Lefschetz periods for the singularity of type $A_N$, we construct a projective representation of the Lie algebra of differential operators on the circle with central charge $h:=N+1$. We prove that the total descendant potential $\D_{A_N}$ of $A_N$-singularity is a highest weight vector. It is known that  $\D_{A_N}$ can be interpreted as a generating function of a certain class of intersection numbers on the moduli space of $h$-spin curves. In this settings our constraints provide a complete set of recursion relations between the intersection numbers. Our methods are based entirely on the symplectic loop space formalism of A. Givental and therefore they can be applied to the mirror models of symplectic manifolds.

\end{abstract}

\maketitle

\sectionnew{Introduction}

It was conjectured by E. Witten \cite{W1} and proved by M. Kontsevich \cite{Ko} that the stable intersection theory on the moduli space $\overline{\M}_{g,n}$ of Riemann surfaces is governed by a unique solution of the KdV hierarchy. Following A. Givental \cite{G3}, we denote this solution by $\D_{\rm pt}$ and we refer to it as the {\em Witten--Kontsevich $\tau$-function}.  More generally, given a compact K\"ahler manifold $X$, let $\overline{\M}_{g,n}(X,d)$ be the moduli space of equivalence classes of degree-$d$ stable holomorphic maps with values in $X$, whose domain is a genus-$g$ Riemann surface equipped with $n$ marked points. Similarly to $\overline{\M}_{g,n}$, there are intersection numbers in $\overline{\M}_{g,n}(X,d)$, called {\em Gromov--Witten invariants} \cite{BF}, \cite{BM}, \cite{Ko2}, \cite{LT}, \cite{S}. One can organize them in a generating function $\D_X$, similar to $\D_{\rm pt}$. It is natural to ask whether $\D_X$ can be uniquely identified with the solution of some integrable hierarchy. In general, it is very hard to approach this question. However, there is a class of manifolds for which the problem looks manageable. To specify them, one has to introduce the notion of {\em Frobenius structure} on a vector space $H$. It consists of a family of Frobenius algebra structures -- one on each tangent space $T_tH,\ t\in H$ -- satisfying certain integrability conditions (see \cite{D}). The Frobenius structure is called {\em semi-simple} if the multiplication in $T_tH$ is semi-simple for generic $t\in H$. The genus-0 Gromov--Witten invariants of $X$ give rise to a Frobenius structure on $H^*(X)$. In case it is semi-simple, Givental conjectured (see \cite{G1}) that $\D_{X}$ is given by a closed formula, which depends only on the semi-simple Frobenius structure and the Witten-Kontsevich $\tau$-function $\D_{\rm pt}.$  Givental's formula was recently proved by C. Teleman \cite{Te}. Since the formula for $\D_X$ makes sense for any semi-simple Frobenius structure, it is natural to investigate a more general problem. Namely, what is the connection between integrable hierarchies and semi-simple Frobenius structures -- see \cite{DLZ} and \cite{DZ}. 

In this article we pursue a different direction, which in some sense is parallel to the above discussion. To begin with, let us recall that there is a different way to characterize $\D_{\rm pt}$. According to V. Kac and A. Schwartz \cite{KacS}, $\D_{\rm pt}$ is a highest weight vector of the Virasoro algebra, which is a central extension of the Lie algebra of vector fields on the circle. Combinatorially, the meaning of the Virasoro constraints is the following. The intersection numbers are obtained by integrating over $\overline{\M}_{g,n}$ certain monomial expressions of the cohomology classes $\psi_1,\ldots, \psi_n$, where $\psi_i$ is the first Chern class of the line bundle on $\overline{\M}_{g,n}$ formed by the cotangent lines at the $i$-th marked point. The Virasoro constraints give rise to a rule for removing the powers of $\psi_n$ and thus they express each intersection number in terms of simpler ones, depending on fewer $\psi$'s or lower genus. The Gromov--Witten invariants are obtained by integrating over $\overline{\M}_{g,n}(X,d)$ monomial expressions in cohomology classes of the type ${\rm ev}_i^*(\phi)\psi_i^k$, where ${\rm ev}_i$ is the evaluation map at the $i$-th marked point and $\phi\in H^*(X)$. One of the fundamental questions in Gromov--Witten theory, which is still open for manifolds $X$ whose quantum cohomology is not semi-simple, is whether $\D_X$ satisfies the Virasoro constraints (see \cite{EHX, EJX, EX} and also \cite{DZ2} and \cite{G3}). They could also be interpreted as rules for removing ${\rm ev}_n^*({\bf 1})\psi_n^k$, where ${\bf 1}\in H^*(X)$ is the unity. More generally, one could ask whether there are rules for removing all cohomology classes ${\rm ev}_n^*(\phi)\psi_n^k,\ \phi\in H^*(X)$, or in terms of generating functions, is it possible to prove that $\D_X$ is a highest weight vector for an algebra larger than the Virasoro algebra? 
   
The above question makes sense for any $\D_X$ arising from a semi-simple Frobenius structure. In \cite{G3}, A. Givental introduced a certain symplectic loop space formalism, which allowed him to prove that his formula satisfies Virasoro constraints. The problem then is to find a larger algebra such that $\D_X$ is still a highest weight vector. There is one case in which the answer is known (see \cite{DV}). Namely, the space of miniversal deformations of an $A_N$-singularity has a semi-simple Frobenius structure and for it $\D_{A_N}$ is given by Givental's formula. In \cite{G1}, it was proved that $\D_{A_N}$ is a solution to the $h$-KdV hierarchy, where $h:=N+1$. In addition, according to the results of \cite{G3}, $\D_{A_N}$ satisfies the {\em string equation}, which is just one of the Virasoro constraints, corresponding to removing ${\bf 1}$ from the intersection numbers. Finally, it was proved in \cite{AM} that a solution to $h$-KdV satisfying the string equation is unique and it is a highest weight vector of the vertex algebra $\W_h$. We are not going to use the theory of vertex algebras here.  The last statement can be reformulated also in the following way (see \cite{FKRW}). The algebra of differential operators on the circle has a unique central extension which is usually denoted by $W_{1+\infty}$. The above statement means that there is a representation of $W_{1+\infty}$ with central charge $h$, such that $\D_{A_N}$ is a highest weight vector. 

In the present article we prove that $\D_{A_N}$ is a highest weight vector for some algebra of differential operators defined in terms of Picard--Lefschetz periods and vertex operators. Our methods are based entirely on the symplectic loop space formalism of A. Givental developed in \cite{G3} and pursued further in \cite{G1} and \cite{GM}. We also prove that after an appropriate change of the variables our constraints coincide with the $\W_h$ constraints. 
 
\subsection{Formulation of the main result.} \label{frobenius}
Let $\T\iso\C^N$ be the space of miniversal deformations of $f(x)={x^{N+1}}/{(N+1)},$ i.e., the points $t=(t^1,\ldots,t^N)\in \T$ parametrize the polynomials
\beq\label{miniversal:deformation}
f_t(x)=\frac{x^{N+1}}{N+1}+t^1 x^{N-1}+\cdots+t^N.
\eeq
To avoid cumbersome notations we put $h:=N+1.$ Each tangent space $T_t\T$ is naturally identified with the algebra of polynomial functions on the critical set ${Crit}\ f_t$: 
$$
\d/\d t^i\in T_t\T\  \mapsto\  \d f_t/\d t^i\, ({\rm mod}\, f_t'(x))\in \C[x]/\langle f_t'(x)\rangle.
$$ 
In particular, the tangent space $T_t\T$ is equipped with an associative, commutative multiplication, which will be denoted by $\bullet_t$. In addition to the multiplication we introduce also a {\em flat} structure on $\T$ via the following residue pairing: 
\beq\label{res}
\( \d/\d t^i\, ,\, \d/\d t^j\)_t:=\sum_{i=1} {\rm res}_{x=\xi_i}
\frac{\d_{t^i}f_t\,\d_{t^j}f_t}{(f_t)'_x}\,\omega,\quad \omega=dx,
\eeq
where $\xi_i,\ 1\leq i\leq N,$ are the critical points of $f_t.$ The flatness here means that we can find a holomorphic coordinate system $(\tau^1,\ldots,\tau^N)$ on $\T,$ in which the above pairing is constant. 

It follows from the definitions that the multiplication $\bullet_t$ is {\em Frobenius} with respect to the residue pairing. It is also known that the following integrability condition holds (for example see \cite{He}): the family of connection operators
\beq\label{connection}
\nabla = \nabla^{\rm L.C.} - \frac{1}{z}\sum_{i=1}^N (\d/\d t^i\bullet_t)dt^i
\eeq
is flat, where $\nabla^{\rm L.C.}$ is the Levi--Cevita connection of the residue pairing. 

Let $H=\C[x]/\langle x^N\rangle$ and $v_i,\ 1\leq i\leq N$ be the basis obtained from the projection of the monomials $x^{N-i}$ on $H$. Put 
\ben
{\bf 0}=(0,\ldots,0,0)\in \T,\quad {\bf 1}=(0,\ldots,0,1)\in \T.
\een 
Let $\tau^i=\tau^i(t)$, $1\leq i\leq N$ be flat coordinates on $\T$ such that $\tau^i({\bf 0})=0$ and the restriction of the coordinate vector field $\d_i:=\d/\d\tau^i$ to the tangent space $T_{\bf 0}\T\iso H$ coincides with $v_i.$ Let us point out that $t^N=\tau^N$, so $\d_N$ is a unity with respect to the Frobenius multiplication $\bullet_t$. Also, the flat structure allows us to identify $\T\iso H$, $t\mapsto \sum_{i=1}^N \tau^i(t)v_i$ and so we may assume that ${\bf 1}=\d_N$. 

\medskip

By definition, the symplectic loop space $\H$ is the space of formal Laurent series in $z^{-1}$ with coefficients in the vector space $H$ (these are series with finitely many positive powers of $z$ and possibly infinitely many negative ones). The space $\H$ is equipped with a symplectic structure:
\ben
\Omega(\phi_1,\phi_2):={\rm res}_{z=0}\(\phi_1(-z),\phi_2(z)\)dz,\quad \phi_1(z),\phi_2(z)\in \H.
\een 
The Darboux coordinate system for $\Omega$ is provided by linear functions $q_k^i$, $p_{k,i}$ defined by the following formula:
\ben
\phi(z) = \sum_{k=0}^\infty q_k^i v_i z^k +  \sum_{k=0}^\infty p_{k,i}\,v^i(-z)^{-k-1}
\een  
where $\{v^i\}_{1\leq i\leq N}$ is a basis of $H$ dual to $\{v_i\}_{1\leq i\leq N}$ with respect to the residue pairing. We also introduce the {\em Fock space}, which by definition is the space of formal power series in the vector variables $q_0,q_1+{\bf 1},q_2,\ldots$, with coefficients in the field $\C_\ge=\C((\ge))$, where $q_k=\sum_i q_k^i v_i$. 

By definition, a {\em vertex operator} is an operator acting on the Fock space of the following type:
\beq\label{vop}
\exp\Big(\sum_{k=0}^\infty (-1)^{k+1}\(I^{(-1-k)},v_i\)\frac{q_k^i}{\ge} \Big)
\exp\Big(\sum_{k=0}^\infty \(I^{(k)},v^i\)\ge\frac{\d}{\d q_k^i}    \Big),
\eeq
where $I^{(n)}\in H.$ In the context of the symplectic loop space formalism the vertex operators will be interpreted as follows. Put $\phi(z)=\sum_{n\in \Z} I^{(n)}(-z)^n\in \H$. The symplectic loop space $\H$ is a direct sum of two Lagrangian subspaces: $\H_-:=H[[z^{-1}]]z^{-1}$ and $\H_+:=H[z].$ We denote by $\phi_+$ (resp. $\phi_-$) the projection of $\phi$ onto $\H_+$ (resp. $\H_-$). The first and second exponent in \eqref{vop} are then quantizations of the linear Hamiltonians $\Omega(\ ,\phi_-)$ and $\Omega(\ ,\phi_+)$ respectively, where the quantization rules are defined by: $\widehat{q}_k^i={q}_k^i/\ge$ and $\widehat{p}_{k,i}=\ge\d/\d{q}_k^i.$  

The period vector $I_a^{(0)}(t,\gl)\in H$, $(t,\gl)\in \T\times \C$ is defined by the following formulas: 
\beq\label{periods}
\(I^{(0)}_a(t,\gl),v_i\):= \int_{a} \d_if_t\frac{\omega}{df_t},\quad 1\leq i\leq N,
\eeq
where $a\in H_0(f_t^{-1}(\gl);\frac{1}{2}\Z)$ is a 1-point cycle and the integral is interpreted as evaluation at $a$ (the coefficients of the cohomology may be chosen even in $\C$, but $\frac{1}{2}\Z$ suffices for our purposes). 

The value of $I_a^{(0)}(t,\gl)$ is well defined only if $(t,\gl)$ is a point outside the {\em discriminant locus} $\{ (t,u)\ |\ u \mbox{ is a critical value of } f_t\},$ and it depends on the choice of a path, avoiding the discriminant, from a fixed reference point in $\T\times \C$ to $(t,\gl)$. It is convenient to choose $({\bf 0},1)$ for a reference point, although any other point outside the discriminant locus would work too.  Furthermore, near $\gl=\infty$, the period vector expands as a Laurent series involving only fractional powers of $\gl$. In particular, it makes sense to put $I^{(n)}_a (t,\gl)=\d_\gl^n I^{(0)}_a (t,\gl),$ where for negative $n$ the operator $\d_\gl^n$ is interpreted as formal integration. 

We are ready to define the vertex operators that will be needed in this article. Let $\Gamma_a(t,\gl,s)$ be the vertex operator corresponding to the series
\ben
\phi_a(t,\gl,s)=\phi_a(t,\gl+s)-\phi_a(t,\gl),\quad \phi_a(t,\gl)=\sum_{n\in \Z}I^{(n)}_a(t,\gl)(-z)^n,
\een
where $s$ is a {\em formal variable}. In other words, our definition should be interpreted as a formal Taylor series in $s:$ 
\ben
\phi_a(t,\gl,s)= \sum_{k= 1}^\infty \Big( \sum_{n\in \Z}I^{(n+k)}_a(t,\gl)(-z)^n\Big)\frac{s^{k}}{k!}.
\een 
Using \eqref{vop}, it is easy to see that the action of the vertex operator $\Gamma_a$ on an element $\D(\q)$ of the Fock space is given by:
\beq\label{vop:action}
\Gamma_a(t,\gl,s)\D(\q) = e^{\frac{1}{\ge}\Omega(\q(z),\phi_-)}\D(\q+\ge \phi_+),
\eeq
where $\phi:=\phi_a(t,\gl,s).$

Finally, we need also the so called {\em phase form}:
\beq\label{phase:form}
\W_{a,b} = \Big(I^{(0)}_a(t,s)-I^{(0)}_a(t,0)\Big)\bullet_t \Big(I^{(0)}_b(t,s)-I^{(0)}_b(t,0)\Big). 
\eeq 
Here 
\ben
I^{(0)}_a(t,s)-I^{(0)}_a(t,0) = \sum_{k= 1}^\infty I^{(k)}_a(t,0)s^k/k!
\een
is interpreted via the Taylor's formula as a formal power series in $s$. Each product $I^{(k)}_a(t,0)\bullet_t  I^{(l)}_b(t,0)$ is a vector in $H$, which should be identified with $T_t\T$ and then (via the residue pairing) with the co-tangent space $T_t^*\T$. So $\W_{a,b} $ is a formal power series in $s$ whose coefficients are 1-forms on $\T$. Notice that the phase form is multi-valued and has poles.  

We say that a function $\D$ from the Fock space satisfies $\W_{A_N}$-constraints if the expression
\beq\label{A_N:constraints}
\sum_{a\in f_{\bf 0}^{-1}(1)} c_a(0,\gl,s)\Gamma_a(0,\gl,s) \, \D\quad 
\eeq
is regular in $\gl.$ Here $c_a(0,\gl,s) = e^{\frac{1}{2}\int_{-{\bf 1}}^{-\gl\,{\bf 1}} \W_{a,a}},$ where the integration path is chosen as follows: first we fix a one point cycle $a_0\in f_{\bf 0}^{-1}(1)$ and then we pick a path from $-{\bf 1}$ to $-\gl{\bf 1}$. For each $a\in f_{\bf 0}^{-1}(1)$, we precompose this path with a closed loop going several times around $0$, such that the parallel transport of $a_0$ is $a$. 

Expression \eqref{A_N:constraints} is independent of the choice of $a_0$ and the path from $-{\bf 1}$ to $-\gl{\bf 1}$, because we can interpret \eqref{A_N:constraints} as the sum over all branches of $c_{a_0}\Gamma_{a_0}\D$. The later means also that  \eqref{A_N:constraints} is invariant under the analytical continuation along a loop around $\gl=\infty$. Therefore, the operator acting on $\D$ expands as a power series in $s$ and a formal series in integral powers of $\gl$: 
\beq\label{A_N:jnk}
\sum_a c_a(0,\gl,s)\Gamma_a(0,\gl,s) = \sum_{k=0}^\infty \sum_{n\in \Z} W_n^k\, \gl^{-n-k}s^k,
\eeq
where $W_n^k $ are some differential operators. The regularity of \eqref{A_N:constraints} means that only non-negative powers of $\gl$ are present, i.e., $W_n^k \D=0$ for $n+k>0.$ 

Our main result is the following theorem.
\begin{theorem}\label{t1}
The total descendant potential $\D_{A_N}$ satisfies the $\W_{A_N}$-constraints.
\end{theorem}

\medskip

 It is easy to obtain the genus-0 limit of the $\W_{A_N}$-constraints. Let $h_a(\p,\q)$ be the linear Hamiltonian $\Omega(\ ,\frac{\d\phi_a}{\d\gl}(0,\gl))$. Then for each $k\geq 2$ we have the following expansion:
\ben
\sum_a c_a(0,\gl,s)\(h_a(\q,\p)\)^k=\sum_{n\in \Z} h_{k,n}(\q,\p)\gl^{-n-k}.
\een
Using the polarization $\H=\H_-\oplus\H_+$ we identify $\H$ with the cotangent bundle. On the other hand, the total descendant potential has the form $\D_{A_N}=e^{\ge^{2g-2}\F^{(g)}}.$ It is known that the graph of the differential $d\F^{(0)}$ is a Lagrangian cone $\L$ in $\H$, which has some very special properties (see \cite{G4} for more details). By taking only the lowest degree terms in $\ge$ in our $\W_{A_N}$-constraints we get:
\begin{corollary}\label{c1} 
The Hamiltonians $h_{k,n}$ vanish for $k\geq 0$, $n>-k$ when restricted to the the Lagrangian cone $\L$.  
\end{corollary}

\medskip

\subsection{$\W_{1+\infty}$-constraints.} 

Let $\Gamma(w,\zeta)$ be a vertex operator, acting on the Fock space $\C_\ge[[t_1,t_2,\ldots]]$, obtained from the composition of the vertex operator 
\beq\label{vop:kp}
  \exp\Big( -\sum_{n=1}^\infty ( w^n-\zeta^n)t_n\Big) \exp\Big(
\sum_{n=1}^\infty \frac{1}{n}(w^{-n}-\zeta^{-n})\d_{t_n}\Big)
\eeq
and the dilaton shift $t_{h+1}\mapsto t_{h+1}-\frac{1}{h+1}.$ We define the differential operators $J_n^k$ by the following Taylor expansion: 
\beq\label{jnk:bosonic}
\sum w^{-N/2}\zeta^{-N/2} (w-\zeta)^{-1}\ \Gamma(w,\zeta)=
\frac{h}{s}+\sum_{k=0}^\infty \sum_{n\in \Z} J_n^k \gl^{-n-k-1}\frac{s^{k}}{k!}
\eeq
where  $w=(h(\gl+s))^{1/h}$, $\zeta=(h\gl)^{1/h}$ and the sum on the left-hand side is over all branches of $\gl^{1/h}$. 

On the other hand, using the change of variables:
\beq\label{change:tq}
t_{-i+kh}= \frac{q_{k-1}^i}{(-i+h)(-i+2h)\cdots (-i+kh)}
\eeq
and the dilaton shift $t_{h+1}\mapsto t_{h+1}-\frac{1}{h+1}$ we identify the Fock space $\C_{\ge}[[q_0,q_1+1,q_2,\ldots]]$ with the subspace of $\C_{\ge}[[t_1,t_2,t_3,\ldots]]$ consisting of all series independent of $t_{h}, t_{2h}, t_{3h}$, etc. Therefore, given an element $\D$ from the Fock space $\C_{\ge}[[q_0,q_1+1,q_2,\ldots]]$ it makes sense to consider the action of $J_n^k$ on $\D.$  
\begin{theorem}\label{t2}
The following statements hold:

\begin{enumerate}
\item[a)] 
The regularity conditions $W_n^k\D = 0$ ($n+k>0, k\geq 0$) are equivalent to $J_n^k\D=0$ ($n+k\geq 0, k\geq 0$).
\item[b)]
The map $-\gl^{n+k}\d_{\gl}^k\mapsto J_n^k$ is a representation of $W_{1+\infty}$ with central charge $h$.
\item[c)] 
If the regularity condition $J_n^k\D=0$, $n+k\geq 0$, holds for $k=0, 1,\ldots h-1$, then it holds for all $k\geq 0.$
\end{enumerate}
\end{theorem}
This theorem will be proved in Section \ref{sec:3}. The proof of a) amounts to changing the variables in \eqref{A_N:jnk} via \eqref{change:tq} and observing that we get \eqref{jnk:bosonic} up to factor, which is invertible and regular in $\gl$. Part c) is a corollary from \cite{FKRW} and b). Finally, to prove b), we construct a representation of $W_{1+\infty}$ (see \cite{vanM}) in the Fermionic Fock space and we prove, using the Boson--Fermion isomorphism, that the representation is the same as the one stated in the lemma.   

\subsection{Higher spin curves}
By definition, an $h$-spin smooth curve is a Riemann surface $\Sigma$ equipped with $n$ marked points $x_1, x_2,\ldots, x_n$, a line bundle $L$ on $\Sigma$, $n$ integer labels $m_1, m_2,\ldots, m_n,$ $0\leq m_i\leq N-1$, and an isomorphism between $L^{\tensor h}$ and the twisted canonical bundle $K\tensor\,\bigotimes_i \O(-m_i x_i)$. The moduli space of equivalence classes of $h$-spin curves is denoted by $\M_{g,n}^{\bf m}$, where ${\bf m}=(m_1,\ldots, m_n)$. It is non-empty iff the degree compatibility condition is met: $h$ is divisible by $2g-2-\sum_i m_i$. In this case the forgetful map $\pi: \M_{g,n}^{\bf m}\rightarrow \M_{g,n}$, which remembers only $\Sigma$ and the marked points, is a degree $h^{2g}$ covering. 

It is known that $\M_{g,n}^{\bf m}$ has a natural compactification $\overline{\M}_{g,n}^{\bf m}$, obtained by analyzing what happens to the sections of the line bundle $L$ when $\Sigma$ acquires a node (see \cite{W1} and \cite{JKV}).  Another important ingredient is the Witten's top Chern class $c_{g,n}^{\bf m}\in H^*(\overline{\M}_{g,n}^{\bf m},\QQ)$. Using the spin structure isomorphism $L^{\tensor h}\iso \omega_\Sigma\tensor\,\bigotimes_i \O(-m_i x_i)$, one can define a map $w:\Omega^{0,0}(L)\rightarrow \Omega^{0,1}(L)$, $s\mapsto \bar\d s + s^N$. This construction, which so far is only over a single point of $\overline{\M}_{g,n}^{\bf m}$ is natural, so $w$ can be interpreted as a map between two bundles over $\overline{\M}_{g,n}^{\bf m}$. Then $c_{g,n}^{\bf m}$ is {\Poincare} dual to the pushforward of the zero locus of $w$ to $\overline{\M}_{g,n}^{\bf m}$. More details can be found in \cite{PV}, \cite{P}, and \cite{W1}.  

Let $H$ be an $N$-dimensional vector space with basis $v_1,\ldots, v_N$ and a non-degenerate bilinear paring defined by $(v_i,v_j)=\delta_{i+j,N+1}$. Put
\ben
\langle v_{i_1}\psi^{k_1},\ldots, v_{i_n}\psi^{k_n}\rangle_{g,n}:=\int_{\overline{\M}_{g,n}^{\bf m} } \psi_1^{k_1}\cdots \psi_n^{k_n}\, c_{g,n}^{\bf m} 
\een
where $\psi_j$ is the first Chern classes of the line bundle on $\overline{\M}_{g,n}^{\bf m}$ formed by the cotangent lines $T^*_{x_j}\Sigma$, and ${\bf m}=(N-i_1,\ldots, N-i_n).$ 

It follows from the work of \cite{JKV}, \cite{P}, \cite{PV}, and \cite{Te} (see also \cite{Sh}), that the total descendant potential of $A_N$-singularity coincides with the following generating function:
\ben
\D_{A_N}=\exp\Big( \sum \frac{1}{n!}\ge^{2g-2} \langle \q(\psi),\ldots, \q(\psi)\rangle_{g,n} \Big),
\een
where the summation is over all $g, n\geq 0$,
$$
\q(\psi) = \sum_{k\geq 0}\sum_{i=1}^N q_k^i v_i \psi^k,
$$ 
and we have to shift $q_1^N\mapsto q_1^N+1$, so that we have a formal series in $q_1^N+1$ and $q_k^i,\ (k,i)\neq (1,N).$ It is easy to see the following. The $J^0$-constraints are empty, because $J^0$ is a constant, the constraint $J^i_{-i-k}\D=0$ comes from a rule for removing $v_{N+1-i}\psi^k$ ($1\leq i\leq N, k\geq 0$), and the rest of the constraints, according to \thref{t2}, are corollaries from the the preceding ones. In particular, the $J^1$-constraints coincide with the Virasoro constraints.

\subsection{Final remark} 
The Virasoro constraints for a point, which in our case correspond to $N=1$, can be proved directly, without using \cite{KacS} and \cite{Ko}. The argument, given by M. Mirzakhani \cite{Mir}, is based on an interpretation of the intersection numbers in terms of symplectic volumes. Then by using the Duistermaat--Heckman formula Mirzakhani obtains some recursion relations, which turn out to be the same as the ones provided by the Virasoro algebra. It would be interesting to see whether our constraints can also be proved geometrically.

\sectionnew{From $\W_{A_N}$ to $\W_{h}$.} \label{sec:3}

The goal in this section is to give a proof of \thref{t2}. 

\subsection{Reduction modulo $h$.}

The change of variables \eqref{change:tq} looks mysterious but it has a very simple purpose: up to terms depending only on $t_h,t_{2h},t_{3h},\ldots $ it just transforms the vertex operator $\Gamma_a(0,\gl,s)$ into $\Gamma(w,\zeta)$ where $w=(h(\gl+s))^{1/h}$ and $\zeta=(h\gl)^{1/h}$ -- see \eqref{vop:kp}. Indeed,  by definition  
\ben
(I^{(0)}_a(0,\gl),v_i)=\int_a {x^{N-i}}/{x^N} = \int_a x^{-i} = (h\gl)^{-i/h}. 
\een
Therefore, for $k\geq 0$ we have: 
\beq\label{period:k}
(I^{(k)}(0,\gl),v_i)=(-1)^ki(i+h)\cdots (i+ (k-1)h)(h\gl)^{-k-i/h},
\eeq
and 
\beq\label{period:k1}
(I^{(-k-1)}(0,\gl),v_i)= \frac{(h\gl)^{k+1-i/h}}{(-i+h)\cdots (-i+(k+1)h)}.
\eeq
Since $(v_i,v_j)=\delta_{i+j,N+1}$, as one can easily verify from \eqref{res}, using our quantization conventions we get:
\ben
 \((I^{(n)}(0,\gl),v_i)v^i(-z)^{n}\)\sphat =
\begin{cases}
-t_{-i+(k+1)h}(h\gl)^{k+1-i/h}   & \mbox{ if } n=-k-1 <0, \\
\frac{1}{i+kh}\,\d_{t_{i+kh}}(h\gl)^{-k-i/h}  & \mbox{ if } n=k \geq 0.
\end{cases}
\een
It follows that the substitution $t_h=t_{2h}=\cdots = 0$ transforms the vertex operator $\Gamma(w,\zeta)$ into $\Gamma_a(0,\gl+s,\gl)$. We call this operator the mod-$h$ reduction of $\Gamma(w,\zeta)$ and denote it by $\leftexp{\rm red}{\Gamma}(w,\zeta).$ 

\subsection{The phase factors.}
Our next goal is to compute the coefficient $c_a(0,\gl,s)=e^{\frac{1}{2}\int_{-{\bf 1}}^{-\gl{\bf 1}} \W_{a,a}}.$ The restriction of the phase form $\W_{a,a}$ to the complex plane in $\T$ spanned by ${\bf 1} $ is: 
\ben
\(I_a^{(0)}(0,-t_N,s),I_a^{(0)}(0,-t_N,s)\)dt_N .
\een
Using the above formula for the period $I_a^{(0)}$ we get:
\ben 
\frac{1}{h}\sum_{i=1}^N 
\Big( (-t_N+s)^{-\frac{i}{h}}-(-t_N)^{-\frac{i}{h}} \Big) 
\Big( (-t_N+s)^{-\frac{h-i}{h}}-(-t_N)^{-\frac{h-i}{h}} \Big).
\een
One ckecks directly that an anti-derivative of this function is
\ben
2\log\ \frac{ \((-t_N+s)h\)^{-\frac{N}{2h}} - \( -t_N   h\)^{-\frac{N}{2h}} }
            { \((-t_N+s)h\)^{\frac{1}{h}} -   \( -t_N   h\)^{\frac{1}{h}}  }.
\een
If we substitute in this formula $t_N=-\gl$ we get $2\log \(\frac{c(s/\gl)}{hs}\)$ where
\beq\label{coeff}
c(s):= \(1+s\)^{-\frac{N}{2h}}  
\frac{s}{\(1+s \)^{\frac{1}{h}} - 1} = 
h+\frac{h^2-1}{24h}s^2 -\frac{h^2-1}{24h}s^3 + O(s^4).
\eeq
Therefore, the coefficient $c_a(0,\gl,s)=e^{\frac{1}{2}\int_{-{\bf 1}}^{-\gl{\bf 1}} \W_{a,a}}$ equals $c(s/\gl)/c(s).$

\medskip

{\em Proof of \thref{t2}, a).}
Recalling the definition \eqref{A_N:jnk} of $W_n^k$  we get:
\ben
c(s)\sum_{k=0}^\infty \sum_{n\in \Z}W_n^k\gl^{-n-k}{s^k}= 
\sum c(s/\gl) \leftexp{\rm red}{\Gamma}(w,\zeta)
\een
where the last sum is over all branches of $\gl^{1/h}$. 

On the other hand, we have $w^{-N/2}\zeta^{-N/2}(w-\zeta)^{-1} = c(s/\gl)/(hs)$. Therefore, equation \eqref{jnk:bosonic} assumes the form:
\ben
\sum c(s/\gl)\Gamma(w,\zeta) = h^2\ + hs\,  \sum_{k=0}^\infty\sum_{n\in \Z} J_n^k\gl^{-n-k-1}\frac{s^k}{k!}. 
\een
Let $\D$ be an element of the Fock space $\C_\ge[[q_0,q_1+{\bf 1}, q_2,\ldots]].$ We need to prove that $\sum c(s/\gl)\Gamma(w,\zeta)\D$ is regular in $\gl$ if and only if $\sum c(s/\gl)\leftexp{\rm red}{\Gamma}(w,\zeta)\D$ is regular in $\gl$. But this is obvious because $ \Gamma(w,\zeta)$ and $\leftexp{\rm red}{\Gamma}(w,\zeta)$ differ by an invertible factor regular in $\gl$, namely $\exp\sum_{k=1}^\infty  t_{kh}\Big((\gl+s)^k-\gl^k\Big)h^k.$ 
\qed

\subsection{The algebra of differential operators on the circle.}

The {\em Fermionic Fock space} $\Lambda^\bullet\(\C[\zeta,\zeta^{-1}]\)$ is a $\Z$-graded vector space, whose degree-$m$ part  $\Lambda^m\(\C[\zeta,\zeta^{-1}]\)$ is spanned by infinite-wedge monomials 
\ben
\zeta^{i_0} \wedge \zeta^{i_1} \wedge \zeta^{i_2} \wedge \ldots
\een
such that $i_0>i_1>i_2>\ldots$ and $i_s=m-s-1$ for $s\gg 0$ (see \cite{KRa}). 
Denote by $\psi_{-i+\frac{1}{2}}$  the operator of wedging by $\zeta^{i}$ and by $\psi_{i-\frac{1}{2}}^*$ the operator of contraction by $\zeta^i$. 

The Lie algebra $gl_\infty$ of all $\Z\times\Z$ matrices having only finitely many non-zero entries can be represented in the Fock space via:
\ben
r(E_{ij})=\psi_{-i+\frac{1}{2}}\psi^*_{j-\frac{1}{2}}
\een
where $E_{ij}\in gl_\infty$ is the matrix defined by $E_{ij}e_k=\delta_{jk}e_i$. This representation can be extended to a {\em projective} representation of $\widetilde{{gl}}_\infty$ -- the Lie algebra of infinite matrices with finitely many non-zero diagonals. Namely, the formulas
\ben
\widehat{r}(E_{ij})=\ \ 
:\psi_{-i+1/2}\psi_{j-1/2}^*:\ \ =
\begin{cases}
\psi_{-i+1/2}\psi_{j-1/2}^* & \mbox{ for } j>0 \\
-\psi_{j-1/2}^*\psi_{i-1/2} & \mbox{ for } j\leq 0.
\end{cases}
\een
define representation of a central extension  $\widehat{{gl}}_\infty= \widetilde{{gl}}_\infty + \C\,c$, with central charge $c=1$ (see \cite{KRa}).

Let $-\gl^{n+k}\d_\gl^k$ ($k\geq 0, n\in \Z$) be a basis of the algebra $w_\infty$ of differential operators on the circle, and let
\ben
e_i = \zeta^{-i-\frac{N}{2}},\quad \zeta =\(h\gl\)^{1/h}.
\een
Then $-\gl^{n+k}\d_\gl^k$ is identified with the infinite matrix:
\beq\label{jnk}
-h^{-n-k}\sum_{i\in \Z}\  \prod_{l=0}^{k-1}\Big(-i-\frac{N}{2}-lh\Big)\
E_{i-nh,i}\ ,
\eeq
and thus we get an embedding of Lie algebras $\phi_{-N/2,h}: w_\infty\hookrightarrow \widetilde{gl}_\infty$ (see \cite{KR, BHY}).  

On the other hand, $w_\infty$ has a unique central extension $W_{1+\infty}=w_\infty+\C\, C$, which can be described as follows. Fix a basis $e_i=\zeta^{-i}$, then each differential operator $-\zeta^{n+k}\d_\zeta^k$ is represented by an infinite matrix, so we have an embedding $\phi_{0,1}:w_\infty \hookrightarrow \widetilde{gl}_\infty.$  The central extension $\widehat{gl}_\infty$ of  $\widetilde{gl}_\infty$ induces a central extension of $w_\infty$, which is isomorphic to $W_{1+\infty}$.  In other words, the map 
\ben
\widehat{\phi}_{0,1}:W_{1+\infty}\rightarrow \widehat{gl}_\infty,\quad 
\zeta^{n+k}\d_\zeta^k\mapsto  \phi_{0,1}(\zeta^{n+k}\d_\zeta^k ),\quad C\mapsto c
\een 
is a Lie algebra embedding. We are going to make use of the following explicit formula for the commutator in $W_{1+\infty}$ (see \cite{KR}):
\beq\label{comm:W}
\left[ \zeta^ke^{xD_\zeta},\zeta^me^{yD_\zeta}\right]_{W_{1+\infty}} = 
\(e^{xm}-e^{yk}\)\zeta^{k+m}e^{(x+y)D_\zeta} + \delta_{k,-m} \frac{e^{xm}-e^{yk}}{1-e^{x+y}}\, C,
\eeq
where $D_\zeta=\zeta\d_\zeta.$

The next lemma is essentially the same as formula (19) in \cite{BHY}. There is, however, a slight difference in the set up here and there, so we will give a separate proof. 
\begin{lemma}\label{extend:phi}
The embedding $\phi_{-N/2, h}: w_\infty\hookrightarrow \widetilde{gl}_\infty$ can be extended to a Lie algebra embedding $\widehat{\phi}_{-N/2, h}: W_{1+\infty}\hookrightarrow \widehat{gl}_\infty$ in the following way: 
\ben
\gl^ne^{xD_\gl} & \mapsto &  
        {\phi}_{-\frac{N}{2},h}\(\gl^ne^{xD_\gl}\) + 
\delta_{n,0}\ \Big(\frac{e^{-\frac{xN}{2h}}}{1-e^{x/h}} - \frac{h}{1-e^x}\Big)\, c \\
C & \mapsto & hc
\een
where $D_{\gl}=\gl\d_{\gl}.$ 
\end{lemma}
\proof
Notice that $\phi_{-N/2, h}=\phi_{0,1}\circ \pi_{-N/2,h}$, where 
\ben
\pi_{-N/2,h}:w_\infty \hookrightarrow w_\infty,\quad 
\gl^kD_\gl^m \mapsto \frac{\zeta^{kh}}{h^k} \Big(\frac{1}{h}(D_\zeta-N/2)\Big)^m.
\een
Therefore, it is enough to construct an extension $\widehat{\pi}_{-N/2,h}:W_{1+\infty}\rightarrow W_{1+\infty}$ of $\pi_{-N/2, h}$.  We are going to look for  $\ga\in w_\infty^*$ and $K\in \C$ such that: 
\ben
\widehat{\pi}_{-N/2,h}(L)={\pi}_{-N/2,h}(L)+ \langle \ga,L\rangle \,C,\quad\widehat{\pi}_{-N/2,h}(C)=K\,C,\quad L\in w_\infty.
\een 
Then
\ben
{\pi}_{-N/2,h}(\gl^ke^{xD_\gl})=
\frac{e^{-\frac{Nx}{2h}}}{h^k}\zeta^{kh}e^{\frac{x}{h}D_\zeta}
\een
and using \eqref{comm:W} it is easy to see that $\widehat{\pi}_{-N/2,h}$ is a Lie algebra homomorphism if and only if 
\ben
\langle \ga, \gl^{k+m}e^{(x+y)D_\gl}\rangle = \delta_{k,-m}\Big(\frac{e^{-(x+y)N/(2h)}}{1-e^{(x+y)/h}} - \frac{K}{1-e^{x+y}}\Big).
\een
Replacing, $k+m$ by $n$ and $x+y$ by $x$, we get:
\beq\label{central:term}
\langle \ga, \gl^{n}e^{xD_\gl}\rangle = \delta_{n,0}\Big(\frac{e^{-\frac{xN}{2h}}}{1-e^{x/h}} - \frac{K}{1-e^{x}}\Big),
\eeq
which equals
\ben
\frac{\delta_{n,0}}{1-e^x}\Big( 
e^{-\frac{xN}{2h}}+ e^{-\frac{x(N-2)}{2h}}+\cdots+e^{-\frac{x(N-2(h-1))}{2h}} - K\Big).
\een
For the RHS in \eqref{central:term} to be a well defined formal power series in $x$, it is necessary and sufficient that $K=h.$
\qed

\subsection{Boson--Fermion isomorphism}
Using the morphism constructed in \leref{extend:phi} and the standard representation $\widehat{r}$, we get a representation of $W_{1+\infty}$ on the Fermionic Fock space with central charge $h$. We would like now to use the Boson--Fermion isomorphism and obtain a representation in the Bosonic Fock space. 

Put 
\ben
\psi(\zeta)=\sum_{i\in \Z}\psi_{i+\frac{1}{2}}\, \zeta^{-i-1}\quad \mbox{ and }\quad
\psi^*(\zeta)=\sum_{j\in \Z}\psi^*_{j+\frac{1}{2}}\, \zeta^{-j-1}.
\een 
The Boson--Fermion isomorphism identifies $\Lambda^m \(\C[\zeta,\zeta^{-1}]\)$ and $\C[[t_1,t_2,\ldots]]q^m$ in such a way that 
\ben
\psi(\zeta)\ \mapsto \ \Gamma_+(\zeta),\quad \psi^*(\zeta)\ \mapsto \ \Gamma_-(\zeta),
\een
where the vertex operators are defined by:
\ben
\Gamma_\pm(\zeta)=q^{\pm 1}\zeta^{\pm m}\exp\Big(\,\pm\, \sum_{n=1}^\infty t_n \zeta^n\ \Big)\exp\Big(\pm \sum_{n=1}^\infty \frac{\d}{\d t_n}\, \frac{\zeta^{-n}}{-n}\Big),
\een
and the following anti-commutation relations are preserved:
\beq\label{comm:1}
[\psi(\zeta),\psi^*(w)]_+ = \delta(\zeta-w)=\sum_{n\in \Z} \zeta^nw^{-n-1},
\eeq
\beq\label{comm:2}
[\psi(\zeta),\psi(w)]_+ = [\psi^*(\zeta),\psi^*(w)]_+=0.
\eeq
It is easy to verify that we have the following Operator Product Expansion:
\ben
-\psi(\zeta)\psi^*(w) = i_{\zeta,w}\frac{1}{w-\zeta} \ \ + \ \  :\psi^*(w)\psi(\zeta):
\een
where $i_{\zeta,w}$ means that we have to expand as a geometric series in the region $|\zeta|>|w|.$

\begin{lemma}\label{t3} The following equality holds: 
\ben
\sum_{n\in \Z}\sum_{k=0}^\infty \widehat{r}\circ \widehat{\phi}_{-N/2,h}(-\gl^{n+k}\d_\gl^k) \gl^{-n-k-1}{s^k}/{k!}\ =\ 
-\frac{h}{s}- \sum_{} w^{-\frac{N}{2}}\zeta^{-\frac{N}{2}}\psi(\zeta)\psi^*(w)
\een
where $w=(h(\gl+s))^{1/h},$  $\zeta=(h\gl)^{1/h},$ each summand on the RHS is interpreted as a formal Taylor series in $s$, and the sum is over all $h$ branches of $\gl^{1/h}.$ 
 \end{lemma}
\proof
First, we prove that:
\beq\label{jnk:2}
\sum_{n\in \Z}\sum_{k=0}^\infty \widehat{r}\circ \phi_{-N/2,h}(-\gl^{n+k}\d_\gl^k) \gl^{-n-k-1}{s^k}/{k!}\ =\ \sum_{} :w^{-\frac{N}{2}}\psi^*(w)\zeta^{-\frac{N}{2}}\psi(\zeta):
\eeq
Using Taylor's formula we get:
\ben
(\gl+{s})^{-\frac{N}{2h}}\psi^*((\gl+s)^{1/h}) = \sum_{k\geq 0}\ \sum_{i\in \Z}\ \d_\gl^k\(\gl^{-\frac{N}{2h}-\frac{i}{h}}\)\psi^*_{i-\frac{1}{2}}\, \frac{s^k}{k!}.
\een
Differentiating with respect to $\gl$, then multiplying by the series 
\ben
\gl^{-\frac{N}{2h}}\psi(\gl^{1/h}) = 
\sum_{j\in \Z}\ \gl^{-\frac{N}{2h}-\frac{j}{N+1}} \psi_{j-\frac{1}{2}},
\een
and rescaling $\gl$ and $s$ by $h$ we get that the RHS in \eqref{jnk:2} equals the sum (over all $k\geq 0$, $i,j\in \Z$) of the following terms:
\beq\label{t3:1}
\prod_{l=0}^{k-1} \Big(-i-\frac{N}{2}-lh\Big)\ (h\gl)^{-\frac{i+j-1}{h} - k-1}\, :\psi^*_{i-\frac{1}{2}}\psi_{j-\frac{1}{2}}: \frac{s^k}{k!}.
\eeq
Note that averaging a formal series in $\gl^{\pm1/h}$ over all branches of $\gl^{1/h}$ kills all fractional powers and leaves the integral ones unchanged. Therefore if we sum \eqref{t3:1} over all branches of $\gl^{1/h}$, then we get  a non-zero answer only for $k\geq 0$, $i\in \Z$ and $j=nh-i+1$. Notice that under the above conditions $\eqref{t3:1}$ is independent of the branch and that  
\ben
:\psi^*_{i-\frac{1}{2}}\psi_{j-\frac{1}{2}}: \ =\  :\psi^*_{i-\frac{1}{2}}\psi_{-i+nh+\frac{1}{2}}:\ =\ -:\psi_{-i+nh+\frac{1}{2}}\psi^*_{i-\frac{1}{2}}:
\een
By definition, the above operator is $-\widehat{r}(E_{i-nh,i})$. Therefore, the coefficient in front of $\gl^{-n-k-1}\frac{s^k}{k!}$ in \eqref{t3:1} coincides with \eqref{jnk}. Notice that the additional factor of $h$ comes from the summation over all branches of  $\gl^{\pm1/h}$. 

The next step will be to use \leref{extend:phi}. In order to do this, we notice that $\sum_{k\geq 0} \gl^{k}\d_\gl^k s^k/k! = e^{xD_\gl}$ where $1+s=e^x$. Indeed, $\gl^{k+1}\d_\gl^{k+1}=\gl^k D\d_\gl^k = (D-k)\gl^k\d_\gl^k$, so if we denote the LHS by $F(x,D)$ where $s=e^x-1$, then it is easy to check that $\d_x F = D F$ and since $F(0,D)=1$ the identity follows.    

So in \eqref{jnk:2} if we replace $\phi_{-N/2,h}$ on the LHS by $\widehat{\phi}_{-N/2,h}$ then according to \leref{extend:phi} we have to add to the RHS the following expression:
\ben
-\frac{1}{\gl} \Big(\frac{e^{-\frac{xN}{2h}}}{1-e^{x/h}} - \frac{h}{1-e^x}\Big)  
\een   
where $1+s/\gl = e^x$. Recall that $w=(h(\gl+s))^{1/h}$ and $\zeta = (h\gl)^{1/h}$, so the above expression is equal to:
\ben
-\frac{h}{s} + h \frac{w^{-N/2}\zeta^{-N/2}}{w-\zeta} = -\frac{h}{s}+\sum \frac{w^{-N/2}\zeta^{-N/2}}{w-\zeta}
\een
where the sum is over all branches of $\gl^{1/h}$. It remains only to notice that
\ben
:\psi^*(w)\psi(\zeta): \ +\  \iota_{\zeta,w}\, \frac{1}{w-\zeta} \  =\  
-\psi(\zeta) \psi^*(w)
\een
where $\iota_{\zeta,w}$ means that we have to expand in the region $|\zeta|>|w|.$
\qed

{\em Proof of \thref{t2}, b)}
According to the Boson--Fermion isomorphism, the operator $-\psi(\zeta)\psi^*(w)$ is transformed into
\ben
-\Gamma_+(\zeta)\Gamma_-(w) = -\zeta^{-1}e^{\sum_{n=1}^\infty t_n \zeta^n} e^{-\sum_{n=1}^\infty \zeta^{-n}/n\d_{t_n}}e^{-\sum_{n=1}^\infty t_n w^n} e^{\sum_{n=1}^\infty w^{-n}/n\d_{t_n}}.
\een 
Given two operators $A$ and $B$ such that $[A,B]=AB-BA$ commutes with both $A$ and $B$, we have $e^Ae^B =e^{[A,B]}e^Be^A$. Applying this for $A$ the translation term of $\Gamma_+$ and $B$ the multiplication term of $\Gamma_-$, we get:
\ben
e^{[A,B]}= \exp\ \sum_{n=1}^\infty (w/\zeta)^n/n = \exp\Big( \log \frac{1}{1-w/\zeta}\Big) = \frac{1}{1-w/\zeta}
\een 
and so 
\ben
-\frac{h}{s}-{w^{-N/2}\zeta^{-N/2}}\psi(\zeta)\psi^*(w) =-\frac{h}{s}+ \frac{w^{-N/2}\zeta^{-N/2} }{w-\zeta} \Gamma(w,\zeta)
\een
where $\Gamma(w,\zeta)$ is the vertex operator defined in \eqref{vop:kp}. Comparing the above formula with \leref{t3} and formula \eqref{jnk:bosonic} we get that 
$$
\widehat{r}\circ \widehat{\phi}_{-N/2,h} (-\gl^{n+k}\d_\gl^k) = J_n^k.
$$ 
Notice that even if we perform the dilaton shift in $J_n^k$, the commutation relations do not change, so we still have a representation of $W_{1+\infty}$ with central charge $h$. 
\qed
 
\subsection{Example.}
We compute explicitly $\leftexp{\rm red}{J}^1(\gl)$, where the left superscript means that we set $t_h=t_{2h}=\cdots = 0$. We do not incorporate the dilaton shift in our computation for typographical reasons. The reader interested in the applications of $J^1$ to higher spin curves should dilaton-shift $t_{h+1}\mapsto t_{h+1}-\frac{1}{h+1}$ our final answer. 

Put $E=\Z\,\backslash \,h\Z$ and denote by $J_m$ the multiplication operator  $-mt_{-m}$ for $m<0$ and the differential operator $\d/\d t_m$ for $m>0$. Then we have: $\phi_a(\gl) = -\sum_{m\in E} J_m \, (h\gl)^{m/h}/m,$ where $a$ corresponds to a choice of $h$-th root of $\gl$. Using the Taylors formula, we get:
\ben
\sum_a \leftexp{\rm red}{\Gamma}(w,\zeta) = h + \sum_a :(\d_\gl\phi_a)^2:\frac{s^2}{2!} + O(s^3)
\een
Since $w^{-N/2}\zeta^{-N/2}(w-\zeta)^{-1} = c(s/\gl)/hs$, using the expansion in \eqref{coeff}, we get
\ben
w^{-N/2}\zeta^{-N/2}(w-\zeta)^{-1} = \frac{1}{hs}\Big(h+\frac{h^2-1}{24h}\gl^{-2} s^2  + O(s^3)\Big)
\een
Substituting in \eqref{jnk:bosonic} we get:
\ben
\leftexp{\rm red}{J}^1(\gl)=\frac{1}{2}\sum_a :(\d_\gl\phi_a)^2: + \frac{h^2-1}{24h}\, \gl^{-2}
\een
or in components:
\beq\label{Virasoro}
\leftexp{\rm red}{J}^1_n = \frac{h^{-n-1}}{2}\sum_{m\in E} :J_mJ_{nh-m}: + \delta_{n,0}\frac{h^2-1}{24h}.
\eeq

\subsection{$\W_{A_1}$ and Virasoro constraints} 
Assume now that $N=1$ and so $h=2$. By definition, the Witten--Kontsevich tau-function is the following generating series:
\beq\label{D:pt}
\D_{\rm pt}=\exp\Big( \sum_{g,n}\frac{1}{n!}\ge^{2g-2}\int_{\overline{\M}_{g,n}}\prod_{i=1}^n (\q(\psi_i)+\psi_i)\Big),
\eeq
where $\q(\psi)=\sum_k q_k \psi^k,$ $(q_0,q_1,\ldots)$ are formal variables, $\psi_i$ ($1\leq i\leq n$) are the first Chern classes of the cotangent-line bundles on $\overline{\M}_{g,n},$ and we have to expand in the powers of $q_0,q_1+1,q_2,\ldots$. 

The substitution \eqref{change:tq}, together with the dilaton shift, gives us $q_k+\delta_{k,1} = (2k+1)!!t_{2k+1}$. According to the main result in \cite{KacS}, $\D_{\rm pt}$ is a highest weight vector for the Virasoro algebra, where the Virasoro operators are $L_n=h^n( \leftexp{\rm red}{J}^1_n).$ On the other hand, it is very easy to see that if $\D$ is a series independent of $t_h,t_{2h},\ldots$ then $\leftexp{\rm red}{J}^1_n\D = J^1_n\D$, so \thref{t2}, c) and a) imply
\begin{corollary}\label{W:A1}
$\D_{\rm pt}$ satisfies the $\W_{A_1}$-constraints.
\end{corollary}

\sectionnew{The total descendant potential $\D_{A_N}$}\label{sec:2}

The goal here is to define $\D_{A_N}$ following A. Givental \cite{G1}. The starting point is a system of differential equations, corresponding to the flat connection \eqref{connection} (see also \eqref{nabla:z} below). The connection has two singularities and we construct two fundamental solutions -- one near each singularity. It turns out that such solutions are symplectic transformations in $\H$ and we can quantize them. This way we obtain certain differential operators, which are applied to a product of $N$ copies of the Witten--Kontsevich tau-function $\D_{\rm pt}.$

\subsection{The system of Frobenius differential equations}\label{SR}
Let $E$ be the vector field on $\T$, whose value at a point $t\in \T$ is $[f_t]\in \C[x]/\langle \d_x f_t\rangle \iso T_t\T$, i.e., 
\ben
E=\sum_{i=1}^N \, \frac{1}{h}\,(i+1)t^i\frac{\d}{\d t^i}.
\een 
Notice that if we assign to $x$ and $t^i$ ($1\leq i\leq N$) degrees $1/h$ and $(i+1)/h$, then $f_t(x)$ is homogeneous of degree 1. Moreover, the residue pairing and the structure constants of the multiplication $\bullet_t$ are also homogeneous. 

The various homogeneity properties can be expressed in terms of the connection operator \eqref{connection}. Namely, $\nabla$ can  be extended to a flat connection on $\T\times \C^*$ in the following way:
\beq\label{nabla:z}
\nabla_{\d/\d z} = \frac{\d}{\d z} -z^{-1}\,\mu+z^{-2} \,E\bullet\ , 
\eeq
where the linear operator $\mu: T_t\T\rightarrow T_t\T$ is defined by 
\ben
\mu(\d/\d t^i) = (i/h - 1/2)\d/\d t^i\quad (1\leq i\leq N)
\een
and $E\bullet$ is the operator of multiplication by the Euler vector field $E$.  The connection $\nabla$ acts on the sections of the pullback bundle $\pi^*T\T,$ where $\pi: \T\times \C^*\rightarrow \T$ is the projection map. 

The connection $\nabla$ is gauge equivalent to $d-(\mu/z) dz$ in a neighborhood of $z=\infty.$ There exists a unique gauge transformation $S_t$, which has the form 
$$
S_t(z)=1+S_1(t)z^{-1}+S_2(t)z^{-k}+\cdots, 
$$
where $S_k(t)$ are linear transformations of $H\iso T_t\T$. Equivalently, $S_t$ is the unique solution to the following system of differential equations:
\beq\label{de_S}
z\d_i S_t = v_i\bullet\, S_t\,, \quad (z\d_z + E)S_t = [\mu,S_t].
\eeq
It is easy to see that $S_t$ satisfies also the initial condition $S_{\bf 0}=1.$

\medskip

Near $z=0$ the system of ordinary differential equations $\nabla J=0$ (see \eqref{de_R} below) has a formal solution of the type $\Psi_tR_te^{U_t/z}$, where the notations are as follows. Let $u^i(t)$ ($1\leq i\leq N$) be the critical values of $f_t$. It is known that for a generic $t$ they form a local coordinate system on $\T$ in which the Frobenius multiplication and the residue pairing are diagonal. Namely,
\ben
\d/\d u^i \, \bullet_t\, \d/\d u^j = \delta_{ij}\d/\d u^j,\quad 
\(\d/\d u^i,\d/\d u^j \) = \frac{\delta_{ij}}{\Delta_i},
\een
where $\Delta_i$ is the Hessian of $f_t$ at the critical point corresponding to the critical value $u^i.$ We denote by $\Psi_t$ the following linear isomorphism:
\ben
\Psi_t:\C^N\rightarrow T_t\T ,\quad e_i\mapsto \sqrt{\Delta_i}\d/\d u^i.
\een
Note that $\Psi_t$ identifies the residue pairing on $T_t\T$ with the standard Euclidean pairing $(e_i,e_j)=\delta_{ij}$. We let $U_t$ be a diagonal matrix with entries $u^1(t),\ldots, u^N(t)$. The series 
$$
R_t(z)=1+R_1(t)z+R_2(t)z^2+\ldots,
$$ 
where $R_k$ are linear operators in $\C^N$, is uniquely determined by the following differential equations:
\beq\label{de_R}
z\d_i \(\Psi R e^{U/z}\) = v_i\bullet_t\, \(\Psi R e^{U/z}\),\quad 
(z\d_z+E)\(\Psi R e^{U/z}\) =\mu\,\(\Psi R e^{U/z}\).
\eeq

\subsection{Quantization of symplectic transformations}

By definition, the {\em twisted loop group} $L^{(2)}GL(H)$ is the group of all symplectic transformations of $\H$ of the type: $M(z)=\sum_k M_k z^k$, where $M_k$ are linear operators in $H$, $z^k$ acts on $\H$ by multiplication, and the sum is over finitely many $k$. It follows from the definition of the symplectic structure $\Omega$ that $M(z)$ is symplectic if and only if $M^T(-z)M(z)=1$, where the transposition is with respect to the residue pairing on $H.$ 

It is known that both series $S_t$ and $R_t$ (see Subsection \ref{SR}) are symplectic transformations of $\H$. Notice that $S_t$ and $R_t$ have the form $e^{A(z)},$ where $A(z)$ is an infinitesimal symplectic transformation. On the other hand, a linear transformation $A(z)$ is infinitesimal symplectic if and only if the map $\f\in \H \mapsto A\f\in \H$ is a Hamiltonian vector field with Hamiltonian given by the quadratic function $h_A(\f) = \frac{1}{2}\Omega(A\f,\f)$. By definition, the quantization of $e^A$ is given by the differential operator $e^{\widehat{h}_A},$ where the quadratic Hamiltonians are quantized according to the following rules:
\ben
(p_{k,i}p_{l,j})\sphat = \ge^2\frac{\d^2}{\d q_k^i\d q_l^j},\quad 
(p_{k,i}q_l^j)\sphat = (q_l^jp_{k,i})\sphat = q_l^j\frac{\d}{\d q_k^i},\quad
(q_k^iq_l^j)\sphat =q_k^iq_l^j/\ge^2.  
\een  
By linearity we obtain a {\em projective} representation of the Poisson Lie algebra of quadratic Hamiltonians of $\H$ on the Fock space. Namely,
\ben
\{F,G\}\sphat = [\widehat{F},\widehat{G}]+C(F,G),
\een
where the cocycle $C$ is $0$ on all pairs of quadratic Darboux monomials except for
\ben
C(p_\ga p_\gb,q_\ga q_\gb)=
\begin{cases}
1, & \mbox{ if } \ga\neq\gb\\
2, & \mbox{ otherwise }
\end{cases},\quad \ga=(k,i),\quad \gb=(l,j).
\een

The action of $\widehat{S}_t$ on an element $F(\q)$ of the Fock space $\C_\ge[[q_0,q_1+{\bf 1},q_2,\ldots]]$ is given by the following formula (see \cite{G3}):
\beq\label{S:fock}
\widehat{S}_t^{-1}\ F(\q)=e^{\frac{1}{2\ge^2}W_t(\q,\q)}F([S_t\q]_+),
\eeq
where the quadratic form is defined by:
\beq\label{W}
W_t(\q,\q)=\sum_{k,l} (v_j,W_{kl}v_i)q_l^iq_k^j,\quad
\sum_{k,l} W_{kl}w^{-k}z^{-l}=\frac{S^T_t(w)S_t(z)-1}{z^{-1}+w^{-1}},
\eeq
and $[\ ]_+$ means truncation of all negative powers of $z$. Similarly (see \cite{G3}),
\beq\label{R:fock}
\widehat{R}_t^{-1}\  F(\q) = \big(e^{\frac{\ge^2}{2}V_t} F\big)(R_t\q), 
\eeq
where $V_t$ is a second order differential operator defined by:
\beq\label{V}
 \sum (v^i, V_{kl} v^j)\frac{\d^2}{\d q_k^i\d q_l^j},\quad 
 \sum_{k,l}V_{kl}w^kz^l = \frac{1-R_t(w)R^T_t(z)}{z+w}.
\eeq
Notice that on the RHS in formula \eqref{R:fock} we first apply the differential operator $e^{\frac{\ge^2}{2}V_t}$ to $F(\q)$ and then we make the substitution $\q\mapsto R_t\q.$ 

\subsection{The total descendant potential} 
Let $t\in \T$ be a semisimple point, i.e., such that the critical values $u^i(t)$ ($1\leq i\leq N$) form a coordinate system. We denote by $\tau=(\tau^1,\ldots,\tau^N)$ the flat coordinates of $t$. The {\em total descendant potential} of $A_N$-singularity is defined in a formal neighborhood of the point $\tau-{\bf 1}z\in \H_+$ by the following formula:
\beq\label{DAN}
\D_{A_N}(\q)= e^{F^{(1)}(t)}\, \widehat{S}^{-1}_t\,\widehat{\Psi}_t\,\widehat{R}_t\,e^{\widehat{U_t/z}}\,\prod_{i=1}^N \D_{\rm pt}(\ge\,\sqrt{\Delta_i}; Q^i\sqrt{\Delta_i}),
\eeq
where: 
\begin{enumerate}
\item[--] 
$Q^i=(Q_0^i,Q_1^i,\ldots )$, $1\leq i\leq N$, are $N$ sequences of variables. 
\item[--]
The formal series $\D_{\rm pt}(\ge\,\sqrt{\Delta_i}\,; Q^i\sqrt{\Delta_i})$ is obtained from the total descendant potential of a point \eqref{D:pt} via the dilaton shift: $\t(z)=Q^i(z)+z$ and rescaling of $\ge$ and $Q^i$ by $\sqrt{\Delta_i}$. 
\item[--] 
$(\widehat{\Psi}_tF)(\q)=F(\Psi^{-1}\q),$ i.e., this is simply the change of variables:
\ben
Q_k^j \sqrt{\Delta_j}=  \sum_{i=1}^N\,\frac{\d u^j}{\d \tau^i}\, q_k^i\quad 
(k\geq 0, 1\leq j\leq N).
\een
\item[--]
The genus-1 potential $F^{(1)}(t)$ is defined in such a way that the RHS of \eqref{DAN} is independent of $t$. The precise value is irrelevant for our purposes.
\end{enumerate}

The product in \eqref{DAN} is a formal power series with coefficients in $\C_\ge$ in the following variables:
\beq\label{Q}
Q_0^i\sqrt{\Delta_i}, Q_1^i\sqrt{\Delta_i}+1, Q_2^i\sqrt{\Delta_i}, \ldots (1\leq j\leq N).
\eeq
The operator $e^{\widehat{U_t/z}}$ is redundant, because its exponent $\widehat{U_t/z}$ is known to annihilate the product of the Witten--Kontsevich $\tau$-functions. The action of the operator $\widehat{R}_t^{-1}$ is given by formula \eqref{R:fock}, where instead of $\q=\sum_{k,i}q_k^i v_i z^k$ one has to use ${\bf Q}=\sum_{k,i} Q_k^i e_i z^k$. In particular,  $\widehat{R}_t^{-1}$ preserves the space of formal power series in the variables \eqref{Q}. 

Furthermore, we have ${\bf 1} = \sum_i ({\bf 1},v^i)v_i$ and 
\ben
Q_1^j\sqrt{\Delta_j}+1 = \sum_{i=1}^N \frac{\d u^j}{\d \tau^i}\, (q_1^i+({\bf 1},v^i) ) - \sum_{i=1}^N\frac{\d u^j}{\d \tau^i}({\bf 1},v^i) + 1.
\een
The second sum is equal to $1$, because the unity in $T_t\T$ is $\sum_i \d/\d u^i$. Therefore, the change of variables $\widehat{\Psi}_t$ is an identification between the Fock space $\C_\ge[[q_0,q_1+{\bf 1}, q_2,\ldots]]$ and the space of formal series in the variables \eqref{Q}.

Finally, by using formula \eqref{S:fock}, it is easy to see that $\widehat{S}_t^{-1}$ is a map from the Fock space $\C_\ge[[q_0,q_1+{\bf 1}, q_2,\ldots]]$ to the space of formal series in $q_0-\tau,q_1+{\bf 1}, q_2,\ldots$. Here one needs to use that $S_1{\bf 1}=\tau$. This follows from the differential equations \eqref{de_S}, which imply $\d_i S_1{\bf 1} = v_i$, and the initial condition $S_{\bf 0}=1$ which implies $S_1({\bf 0})=0$. 

More details can be found in \cite{G1}.

\sectionnew{Symplectic Action on Vertex Operators}

Let $S=1+S_1 z^{-1}+S_2 z^{-2}+\cdots$ and $R=1+R_1z+R_2z^2+\cdots$ be two symplectic transformations of $\H$. The adjoint action of their quantizations $\widehat{S}$ and $\widehat{R}$ on a vertex operator of the type $e^{\widehat{\phi}(\gl,s)}$, where $\phi(\gl,s)=\phi(\gl+s)-\phi(\gl)$ is given by the following formulas (see \cite{G2}):
\beq\label{S}
\widehat{S} \,e^{\widehat{\phi}(\gl,s)}\, \widehat{S}^{-1} = e^{W(\phi(\gl,s)_+,\phi(\gl,s)_+)/2} e^{(S\phi(\gl,s))\sphat},
\eeq
where $W$ is the quadratic form defined by \eqref{W}, and
\beq\label{R}
\widehat{R}^{-1} \,e^{\widehat{\phi}(\gl,s)}\, \widehat{R} = e^{V\phi(\gl,s)_-^2/2} e^{(R^{-1}\phi(\gl,s))\sphat}, 
\eeq
where $\phi(\gl,s)_-$ is identified with the linear function $\Omega(\phi(\gl,s)_-,\ )$ and $V$ is the second order differential operator defined by \eqref{V}.

\subsection{Remark}
In our settings the exponents of the vertex operators have the form:
\ben
\phi(\gl,s)=\sum_{n\in\Z} \(I^{(n)}(\gl+s)-I^{(n)}(\gl)\)(-z)^n,\quad I^{(n+1)}(\gl)=\d_\gl I^{(n)}(\gl).
\een 
In formula \eqref{S}, the expression $S\phi(\gl,s)$ could be interpreted in the  formal $\gl^{-1}$-adic sense, provided that each period $I^{(n)}(\gl)$ expands as a Laurent series near $\gl=\infty$. Similarly, formula \eqref{R} admits a formal $(\gl-u)$-adic interpretation, provided that $I^{(n)}$ expands as a Laurent series near $\gl=u$ for some $u\in \C.$

\subsection{Symplectic translation}
We recall the fields $\phi_a(t,\gl,s)\in \H$, which were introduced in the introduction. They satisfy the following differential equations: 
\beq\label{picard:fuchs}
z\d_i\, \phi_a(t,\gl,s) = v_i\bullet\, \phi_a(t,\gl,s),\ 1\leq i\leq N, \quad 
z\d_\gl \, \phi_a(t,\gl,s) = \phi_a(t,\gl,s).
\eeq
The last equation is trivial. The rest follow from the fact that the form $\omega=dx$ is {\em primitive} in the sense of K. Saito \cite{SaK} (see also \cite{He}, Chapter 11). An elementary proof, using only Cauchy residue theorem, can be found in \cite{MT}. The argument in \cite{MT} is for the space of miniversal deformations of a Laurent polynomial in one variable, but after a minor modification it works for $A_N$ singularity as well.   

\begin{lemma}\label{S:phi}
The following formula holds: $S_t\phi_a(0,\gl,s)=\phi_a(t,\gl,s).$
\end{lemma}
\proof
Both $S_t\phi(0,\gl,s)$ and $\phi(t,\gl,s)$ satisfy the same ordinary differential equations in $t$ and since $S_{\bf 0}=1$, they satisfy the same initial condition at $t={\bf 0}$. 
 \qed

Let $t\in \T$ be a semisimple point. Let $u^i=u^i(t)$ be one of the critical values. We fix a pair of 1-point cycles $a,b\in H_0(f_t^{-1}(\gl);(1/2)\Z)$ such that $\gb:=(a-b)/2$ is a {\em vanishing cycle}, i.e., $\gb$ vanishes when transported from $(t,\gl)$ to $(t,u^i)$. 

Notice that in the case of $A_1$ singularity $u^i(t)=t$ and
\ben
\phi_\gb (t,\gl,s)=\sum_n (-z\d_\gl)^n \Big(\frac{1}{\sqrt{2(\gl+s-t)}} -\frac{1}{\sqrt{2(\gl-t)}}\Big)(-z)^n.
\een
We will denote the above sum by $\phi_{A_1}(t,\gl,s).$
\begin{lemma}\label{R:phi}
For $\gl$ near $u^i$ the following formula holds: 
\ben
\phi_\gb(t,\gl,s)= \Psi_t R_t \phi_{A_1}(u^i,\gl,s)e_i=\Psi_t R_t e^{U_t/z}\, \phi_{A_1}(0,\gl,s).
\een
\end{lemma}
\proof
According to A. Givental (see  Theorem 3 in \cite{G2}) we have: 
\ben
\sum_{n\in\Z} I^{(n)}_\gb (t,\gl)(-z)^n =\Psi_t R_t 
\sum_{n\in\Z} (-z\d_\gl)^n \frac{1}{\sqrt{2(\gl-u^i)}}\, e_i.
\een
Replacing $\gl$ with $\gl+s$ and then subtracting the above formula we obtain the formula stated in the lemma.
\qed

\subsection{Phase factors and the symplectic structure $\Omega$}

The phase factors in \eqref{S} and \eqref{R} can be expressed in terms of the symplectic structure. 
\begin{lemma}\label{v:omega}
Let us identify the vectors $\f,\overline{\f}\in \H_-$ with linear functions via $\Omega(\f,\ )$ and $\Omega(\overline{\f},\ )$. Then we have 
\ben
V\f\overline{\f} = \Omega\([R^{-1}\f]_+,[R^{-1}\overline{\f}]_-\).
\een
\end{lemma}
\proof
Using the definition of $V_{kl}$ and induction on $l,$ it is easy to prove that
\ben
V_{kl}=(-1)^{l+1}R_{k+l+1}+(-1)^{l}R_{k+l}R_1^T+\ldots+(-1)^{l+1-l}R_{k+1}R_l^T.
\een 
Putting 
$$
\f=\sum_{k\geq 0} f_k(-z)^{-k-1}\quad\mbox{and}\quad
\overline{\f}=\sum_{l\geq 0} \overline{f}_l(-z)^{-l-1}
$$
we obtain
\ben
V\f\overline{\f} = \sum_{k,l} (f_k, V_{kl}\overline{f}_l)= 
\sum_{k,l}\sum_{i=0}^l (-1)^{l+1-i}(R_{k+l+1-i}^Tf_k,R_i^T\overline{f}_l).
\een
The last expression should be compared with 
\ben
{\rm res}_{z=0}\([R^T(z)\f(-z)]_+,[R^T(-z)\overline{\f}(z)]_-\)dz.
\een
The lemma follows, because $R^T(-z)=R^{-1}.$
\qed

The next lemma can be proved by a similar argument. 
\begin{lemma}\label{w:omega}
For $\q,\overline{\q}\in \H_+$, the following formula holds: 
\ben
W(\q,\overline{\q})=\Omega([S\q]_+,[S\overline{\q}]_-).
\een
\end{lemma}

\subsection{The Phase form} 

The symplectic pairing $\Omega(\phi_\ga(t,\gl,s)_+,\phi_\gb(t,\gl',s)_-)$ does not make sense, because the coefficient in front of $s^k$ for a fixed integer $k$ is an infinite sum. However, if we pick $u\in \C\cup\{\infty\}$ and expand the periods $I^{(n)}_\ga(t,\gl)$, $I^{(n)}_\gb(t,\gl')$ ($n\in \Z$) as Laurent series near $\gl=u$ and $\gl'=u$ respectively, then the symplectic pairing determines a well defined element in the following space of formal Laurent series in two variables:
\ben
\C((\gl-u,\gl'-u)):=\C((\gl-u))((\gl'-u))\cap \C((\gl'-u))((\gl-u)),
\een
where $\gl-u$ and $\gl'-u$ should be replaced by $\gl^{-1}$ and $(\gl')^{-1}$ if $u=\infty.$

\begin{lemma}\label{primitive}
Let $\ga$ and $\gb$ be arbitrary cycles. Then 
\ben
d\Omega(\phi_\ga(t,\gl,s)_+,\phi_\gb(t,\gl',s)_-) = I^{(0)}_\ga(t,\gl,s)\bullet I^{(0)}_\gb(t,\gl',s)  ,
\een
where $d$ is the de Rham differential on $\T.$
\end{lemma}
\proof
By definition, $\Omega(\phi_\ga(t,\gl,s)_+,\phi_\gb(t,\gl',s))$ equals
\ben
\sum_{k=0}^\infty (-1)^{k+1}\(I^{(k)}_\ga(t,\gl,s),I^{(-k-1)}_\gb(t,\gl',s)\).
\een
On the other hand, using the differential equations \eqref{picard:fuchs} we get: $$
\d_i I^{(n)}_\ga(t,\gl,s) = -v_i\bullet_t I^{(n+1)}_\ga(t,\gl,s).
$$ 
So if we differentiate with respect to $\tau^i$ we get
\ben
\sum_{k=0}^\infty (-1)^{k}\Big(\(v_i\bullet I^{(k+1)}_\ga(t,\gl,s),I^{(-k-1)}_\gb(t,\gl',s)\) +\(I^{(k)}_\ga(t,\gl,s),v_i\bullet I^{(-k)}_\gb(t,\gl',s)\)\Big) 
.
\een
The only term in the above sum that survives is 
$
\(I^{(0)}_\ga(t,\gl,s),v_i\bullet I^{(0)}_\gb(t,\gl',s)\).
$
\qed

\medskip

Let $t$ be a semi-simple point and $u^i(t)$ be one of the critical values of $f_t$. Given $\gl$ sufficiently close to $u^i$ we pick a pair $a$ and $b$ of one point cycles from $H_{0}(f_t^{-1}(\gl),(1/2)\Z)$ such that $\gb=(a-b)/2$ is a vanishing cycle.  
\begin{lemma}\label{V:bb}
The following formula holds:
\ben
d V_t(\phi_\gb(t,\gl,s))_-^2 = -\W_{\gb,\gb}(t-\gl\,{\bf 1})+ 
\Big(\frac{1}{\sqrt{2(\gl+s-u^i)}}-\frac{1}{\sqrt{2(\gl-u^i)}}\Big)^2du^i,
\een
where $d$ is the de Rham differential on $\T$. 
\end{lemma}
\proof
Put $\f=\Psi^{-1}\phi_\gb(t,\gl,s)$ and $\f'=\Psi^{-1}\phi_\gb(t,\gl',s)$ for brevity. According to \leref{v:omega},  $ V_t\f_-\f'_-$ is equal to:
\ben
\Omega([R_t^{-1}\f_-]_+,[R_t^{-1}\f'_-]_-) 
= \Omega([R_t^{-1}\f_-]_+,[R_t^{-1}\f']_-) =\Omega([R_t^{-1}\f]_+-R_t^{-1}\f_+,R_t^{-1}\f' ).
\een
Therefore, we get
\beq\label{v:omega1}
\Omega([R^{-1}\f]_+,[R^{-1}\f']_-)-\Omega(\f_+,\f'_-).
\eeq
On the other hand (see \leref{R:phi}) 
\ben
R_t^{-1}\f_\gb(t,\gl,s) = \sum_{n} I_{A_1}^{(n)}(u^i,\gl,s)(-z)^n.
\een
For $A_1$ singularity we have
\ben
I^{(0)}_{A_1}(u^i,\gl,s)= 
\Big( \frac{1}{\sqrt{2(\gl+s-u^i)}}-\frac{1}{\sqrt{2(\gl-u^i)}}\Big)
\een
and thus the lemma follows by applying \leref{primitive} and setting $\gl'=\gl$.
\qed

A similar argument yields that (see also Section 7 in \cite{G2})
\ben
dW_t(\phi_a(0,\gl,s)_+,\phi_a(0,\gl,s)_+)= \W_{a,a}(t-\gl{\bf 1}). 
\een 
On the other hand $W_{\bf 0}=0$ because $S_{\bf 0}=1$. Therefore, we get
\beq\label{w:aa}
W_t(\phi_a(0,\gl,s)_+,\phi_a(0,\gl,s)_+)=\int_{-\gl{\bf 1}}^{\tau-\gl{\bf 1}}\W_{a,a}.
\eeq






Finally, in a neighborhood of $\gl=u^i$ we have the following vertex operators factorization (see Proposition 4 in \cite{G2}): 
\ben
\Gamma_a(t,\gl,s) = e^{K_a}\Gamma_\ga(t,\gl,s)\Gamma_\gb(t,\gl,s) 
\een
where $\ga=(a+b)/2$ and 
\ben
K_a=-\Omega\Big( (\phi_\ga(t,\gl,s))_+,(\phi_\gb(t,\gl,s))_-\Big). 
\een
Notice that $\phi_\ga$ is analytic near $\gl=u^i$. This is because each period $I^{(0)}_\ga(t,\gl)$ expands in the powers of $(\gl-u^i)^{1/2}$ and has a pole of order at most 1/2. On the other hand, the cycle $\ga$ is invariant under the local monodromy and so the period $I^{(0)}_\ga(t,\gl)$ must be single-valued near $\gl=u^i$, i.e., the corresponding expansion has only integral powers of $\gl-u^i$.
Now it is easy to see that the symplectic pairing of $\phi_{\ga +}$ and $\phi_{\gb -}$ is well defined in the $(\gl-u^i)$-adic sense. It follows from \leref{primitive}, with $\gl=\gl'$ that $dK_a = -\W_{\ga,\gb}(t-\gl{\bf 1}).$

\subsection{Periods of the phase form}
A crucial step in our proof of \thref{t1} is that certain periods of the phase form vanish. Using the map 
\beq\label{map}
\T\times \C\rightarrow \T,\quad (t,\gl)\mapsto t-\gl{\bf 1}
\eeq
we interpret the integral $\int_\gamma \W_{\ga,\gb}$, where $\gamma$ is a path in $\T\times\C$ and $\ga$, $\gb$ are cycles, as $\int_{\gamma} \widetilde{\W}_{\ga,\gb}$, where $\widetilde{\W}_{\ga,\gb}$ is the pullback via \eqref{map} of $\W_{\ga,\gb}.$   
\begin{lemma}\label{period:beta}
Let $\gamma$ be a small loop around a generic point on the discriminant and $\gb$ be a cycle vanishing at that point. Then $\int_\gamma \W_{\gb,\gb} = 0.$
\end{lemma}
\proof
We may assume that $\gamma$ is in the $\gl$-plane $\{t\}\times \C.$ By definition, the restriction of the pullback via \eqref{map} of $\W_{\gb,\gb}$ to $\gamma$ has the form:
\ben
\W_{\gb,\gb} = -\sum_{n=2}^\infty\Big(\ \sum_{\substack{k\geq 1,l\geq 1\\k+l=n}}^\infty \ \frac{1}{k!l!}\,\(I^{(k)}_\gb(t,\gl), I^{(l)}_\gb(t,\gl)\) \Big)s^n\ d\gl. 
\een
Using that $I^{(k)}_\gb(t,\gl)=\d_{\gl}^kI^{(0)}_\gb(t,\gl)$ and integration by parts we get 
\ben
\int_\gamma \W_{\gb,\gb} = -\sum_{n=2}^\infty\int_\gamma\Big(\ \sum_{\substack{k\geq 1,l\geq 1\\k+l=n}}^\infty \ \frac{(-1)^k}{k!l!}\,\(I^{(0)}_\gb(t,\gl), I^{(k+l)}_\gb(t,\gl)\) \Big)s^n\ d\gl.
\een
On the other hand, 
\ben
0=(1-1)^n=\sum_{k=0}^n {n \choose k}(-1)^k
\een
and thus
\ben
\sum_{k+l=n}\frac{(-1)^k}{k!l!}=-\frac{1}{n!}(1+(-1)^n).
\een 
We get that $n=:2m$ should be even and that the period of the phase form turns into
\ben
\sum_{m=1}^\infty\int_\gamma
\(I^{(0)}_\gb(t,\gl), I^{(2m)}_\gb(t,\gl)\)
\frac{s^{2m}}{(2m)!} \ d\gl = \int_\gamma \(I^{(0)}_\gb(t,\gl), I^{(0)}_\gb(t,\gl,s)+I^{(0)}_\gb(t,\gl,-s)\)d\gl .
\een 
On the other hand, according to \leref{R:phi} we have 
\ben
I^{(0)}_\gb(t,\gl)=\Psi_t\sum_k R_k \d_\gl^{-k} I^{(0)}_{A_1}(u^i,\gl),\quad
 I^{(0)}_\gb(t,\gl,\pm s)=\Psi_t\sum_l R_l \d_\gl^{-l} I^{(0)}_{A_1}(u^i,\pm s, \gl).
\een
However, $\Psi_t$ is an isometry and $R^T_t(-\d_\gl)R_t(\d_\gl)=1$, because $R_t$ is a symplectic transformation. Using these two facts and integration by parts we get that the period equals
\ben
\int_\gamma \(I^{(0)}_{A_1}(u^i,\gl),I^{(0)}_{A_1}(u^i, s, \gl)\)d\gl + 
\int_\gamma \(I^{(0)}_{A_1}(u^i,\gl),I^{(0)}_{A_1}(u^i, -s, \gl)\)d\gl.
\een
Expanding in the powers of $\pm s$ via the Taylor's formula we get
\ben
I^{(0)}_{A_1}(u^i,\pm s, \gl) =  \sum_{k\geq 1} \frac{(\mp s)^k}{k!} (2k-1)!! (2(\gl-u^i))^{-1/2-k}\, e_i.
\een
Therefore,
\ben
\int_\gamma \, \(I^{(0)}_{A_1}(u^i,\gl),I^{(0)}_{A_1}(u^i, \pm s, \gl)\) \, d\gl = 
\sum_{k\geq 1} \frac{(\mp s)^k}{k!} (2k-1)!! \int_\gamma (2(\gl-u^i))^{-1-k} d\gl = 0.
\een
\qed 

Now we are ready to prove the following global vanishing property. 
\begin{lemma}\label{period:alpha}
Let $\gamma$ be any loop in $\T\times \C$ avoiding the discriminant and $\ga$ is a cycle invariant under the parallel transport along $\gamma$. Then $\int_\gamma\W_{\ga,\ga}=0.$
\end{lemma}
\proof
The proof here follows the argument in \cite{G2}, Proposition 1. The parallel transport along a closed loop determines a monodromy transformation in $H_0(f_{\bf 0}^{-1}(1);\C)$. It is not hard to see that all monodromy transformations form a group isomorphic to the quotient of the braid group on $N+1$ strands by the relations: $r_1^2=1,\ldots, r_N^2=1$, where $r_i$ is the braid whose strands are straight except for the ones from $i$ to $i+1$ and from $i+1$ to $i$. In fact, all vanishing cycles form a root system (in $H_0(f_{\bf 0}^{-1}(1);\R)\iso \R^{N+1}$) of type $A_N$ and the monodromy group is isomorphic to $S_{N+1}$ -- the Weyl group of the root system of type $A_N$. 

Now we use the fact that if a vector $\ga\in \R^{N+1}$ is invariant under a permutation $\sigma$ then $\sigma$ can be decomposed into a product of transpositions that leave $\ga$ invariant. This means that our path $\gamma$ can be decomposed into $\gamma_1'\ldots\gamma_k'\gamma_1^2\ldots\gamma_l^2$ where $\gamma_i'$ and $\gamma_j$ are simple loops around the discriminant and the cycle $\ga$ is invariant along $\gamma_i'.$ We have $\int_{\gamma_i'}\W_{\ga,\ga}=0$, because $\ga$ is invariant along $\gamma_i'$, which in particular implies that the periods $I_\ga^{(n)}(t,\gl)$ -- hence the phase form $\W_{\ga,\ga}$ -- are holomorphic. 

The loop $\gamma_j$ goes around a generic point $(t,u^i(t))$ on the discriminant. Let $\gb$ be a vanishing cycle. Put $\ga=\ga'+\langle\ga,\gb\rangle \gb/2,$ where $\ga'$ is invariant along $\gamma_j$ and $\langle\ ,\ \rangle$ is the intersection pairing. Then we have: 
\begin{enumerate}
\item $\int_{\gamma_j^2}\W_{\ga',\ga'} = 0$, because $\W_{\ga',\ga'}$ is holomorphic near $(t,u^i(t))$, 
\item  $\int_{\gamma_j^2}\W_{\ga',\gb} = 0$, because the parallel transport along $\gamma_j$ transforms $\gb$ into $-\gb$. So the period may be written as: 
\ben
\int_{\gamma_j^2}\W_{\ga',\gb} = \int_{\gamma_j}\W_{\ga',\gb} + \int_{\gamma_j}\W_{\ga',-\gb} =0. 
\een
\item $\int_{\gamma_j^2}\W_{\gb,\gb} = 0$, according to \leref{period:beta}.
\end{enumerate}  
It follows that $\int_{\gamma_j^2}\W_{\ga,\ga} = 0.$
\qed

\sectionnew{Proof of \thref{t1}}

We split the proof of \thref{t1} into 3 steps. 

\subsection{From descendants to ancestors.}\label{step1} The following formal series is called the {\em total ancestor potential} of the $A_N$-singularity:  
\ben
\A_t:=\widehat{\Psi}_t\widehat{R}_te^{\widehat{U_t/z}}\,\prod_{i=1}^N\D_{\rm pt}(\ge\sqrt{\Delta_i};Q^i\sqrt{\Delta_i}).
\een 
We have $\D_{A_N}=e^{F^{(1)}(t)}\widehat{S}_t^{-1}\A_t$. Using the conjugation formula \eqref{S}, \leref{S:phi}, and formula \eqref{w:aa} we get that the proof of \thref{t1} amounts to proving that the series
\beq\label{anc:constraints}
\sum_{a} c_a(t,\gl,s)\Gamma_a(t,\gl,s)\A_t
\eeq
is regular in $\gl,$ where $c_a(t,\gl,s)=e^{\int_{-{\bf 1}}^{t-\gl{\bf 1}} \W_{a,a}}.$ The integration path in $c_a$ is the composition of the path from $-{\bf 1}$ to $-\gl{\bf 1}$ used in the definition of $c_a(0,\gl,s)$ and an arbitrary path from $-\gl{\bf 1}$ to $t-\gl{\bf 1}.$ The regularity of \eqref{anc:constraints} is interpreted the same way as that of \eqref{A_N:constraints}.

\subsection{ From regularity at $\gl=\infty$ to regularity at the critical values of $f_t$.} \label{step2}

The total ancestor potential $\A_t$ has the following crucial property. If we write
\ben
\A_t =\exp \sum_{g=0}^\infty \overline{\F}^{(g)}(\q)\ge^{2g-2}\quad \in \quad
\C[[\q,\ge^{\pm 1}]],
\een
then each $\overline{\F}^{(g)}$ satisfies the following $(3g-3+r)$-jet constraints:
\ben
\left. \frac{\d^{r}\F^{(g)}}{\d q_{k_1}^{i_1}\ldots \d q_{k_r}^{i_r}}\right|_{\q(z)=-z} = 0 \quad \mbox{ if } k_1+\ldots + k_r\geq 3g-3+r. 
\een

Notice that 
\ben
\Gamma_a(t,\gl,s)\A_t \quad\in\quad \C[[\q,\ge^{\pm 1},\gl^{\pm 1},s]]. 
\een
Given a multi-index of the type $I=\{(k_1,i_1),\ldots,(k_r,i_r)\}$, we put $\q^I=q_{k_1}^{i_1}\ldots q_{k_r}^{i_r}$ and we say that $I$ has length $l(I):=r.$
\begin{lemma}\label{tameness} The coefficient in front of each $\q^I\,\ge^G\, s^M$ in $\Gamma_a(t,\gl,s)\A_t$ depends polynomially on finitely many periods $I^{(n)}_a(t,\gl)$. 
\end{lemma}
\proof
For brevity, put $\phi=\phi_a(t,\gl,s)$. Using that the action of the vertex operator on the Fock space is given by \eqref{vop:action} we get
\beq\label{vop:anc}
\Gamma_a(t,\gl,s)\A_t=\exp\Big( \frac{1}{\ge}\Omega(\q(z),\phi_-) + \sum_{g=0}^\infty \overline{\F}^{(g)}(\q+\ge \phi_+)\ge^{2g-2}\Big).
\eeq
Expanding the genus-$g$ term in the above sum in the powers of $\phi_+$ we get:
\ben
\sum \frac{1}{|{\rm Aut}|}\ge^{2g-2+r}
\frac{  \d^r\overline{\F}^{(g)}                         }
     {\d q_{k_1}^{i_1}\ldots \d q_{k_r}^{i_r} }(\q)(I_a^{(k_1)},d\tau^{i_1})\ldots (I_a^{(k_r)},d\tau^{i_r}), 
\een
where the sum is over all multi-indexes $\{(k_1,i_1),\ldots, (k_r,i_r)\}$, ordered lexicographically, and $|{\rm Aut}|$ is the corresponding number of index-automorphisms.  Since we are interested in the coefficient in front of $\ge^{G}$ for a fixed $G$ we get that $g\leq G$ and $r\leq G$, i.e., there are only finitely many combinations of genus-$g$, partial derivative of order $r$ terms which contribute to our coefficient. 
On the other hand among all monomials in $\overline{\F}^{(g)}$ only the ones of length less or equal to $r+l(I)$ could contribute and thus due to the $(3g-3+r)$-jet property we have $k_i\leq 3g-3+r+l(I)$. We get that we have finitely many choices for $k_1,\ldots,k_r$. Finally, since 
\ben
I^{(k_i)}_a(t,\gl,s)=\sum_{n\geq 1} I^{(k_i+n)}_a(t,\gl)\frac{s^n}{n!}
\een
and we are interested in the coefficient in front of a fixed $s^M$, we get that the coefficient in front of $\q^I\ge^Gs^M$ in the exponent of \eqref{vop:anc} depends polynomially on the periods $I^{(n)}_a(t,\gl)$. Notice that we also have the following relations: $M+G\geq 0$ and $M>0$. The remaining term $\ge^{-1}\Omega(\q,\phi_-)$ also has this form. Therefore, the exponent in \eqref{vop:anc} has the following property:
\begin{enumerate}
\item[(*)]it is a series of the type $\sum c_{I,G,M}\q^I\ge^G s^M,$ where the sum is over all multi-indexes $I$, integers $M\geq 1$, and integers $G\geq -M$, whose coefficients $c_{I,G,M}$ depend polynomially on finitely many periods $I^{(n)}(t,\gl)$.
\end{enumerate}
It is straightforward to check that property (*) is preserved under exponentiation. The lemma follows.
\qed

Using \leref{tameness}, we get that the regularity condition is equivalent to the polynomiality of certain meromorphic functions, which are given by some polynomials of the periods $I^{(n)}(t,\gl)$. Apriory, such functions are defined and single valued in the whole complex plane except possibly for $\gl=u^i$ ($1\leq i\leq N$) -- the critical values of $f_t$. So we have to prove that for each $\gl=u^i$ the expression \eqref{anc:constraints} has no pole at $\gl=u^i$. 

\subsection{Regularity at the critical values.}\label{step3}

Let $u^i$ be any of the critical values of $f_t$. All one point cycles in the sum \eqref{anc:constraints}, except for two, which will be denoted by $a$ and $b$, are invariant under the local monodromy transformation around the discriminant near the point $(t,u^i)$. This means that all terms in the sum are regular at $\gl=u^i$, except for the two exceptional ones. Therefore, we have to prove that the expression
\beq\label{ui:constraints}
\Big(c_a(t,\gl,s)\Gamma_a(t,\gl,s)+c_b(t,\gl,s)\Gamma_b(t,\gl,s)\Big)\, \A_t
\eeq
expanded as a formal series in $(\gl-u^i)^{\pm 1}$ has no poles. 

Put $\ga=(a+b)/2$ and $\gb=(a-b)/2$. Then we have the following vertex operators factorization (using $a=\ga+\gb$ and $b=\ga-\gb$):
\beq\label{vop:factorization}
\Gamma_a = e^{K}\Gamma_{\ga}\Gamma_\gb,\quad \Gamma_b = e^{-K}\Gamma_\ga\Gamma_{-\gb},\quad K=-\Omega(\phi_\ga(t,\gl,s)_+,\phi_\gb(t,\gl,s)_-).
\eeq
The ancestor potential $\A_t$ has the form $\widehat{\Psi}_t\widehat{R}_t \prod_i \A_i$.  In \eqref{ui:constraints}, we factorize the vertex operators according to \eqref{vop:factorization}. We can drop the vertex operator $\Gamma_\ga$ because it is analytic near $\gl=u^i$. Then after conjugating by $\widehat{\Psi}_t\widehat{R}_t $ and using formula \eqref{R} and \leref{R:phi} we see that the regularity of \eqref{ui:constraints} is equivalent to the regularity of the following expression: 
\ben
\Big(\overline{c}_a(t,\gl,s)\Gamma_{+}(u^i,\gl,s) + 
\overline{c}_b(t,\gl,s)\Gamma_{-}(u^i,\gl,s)\Big) \A_i,
\een 
where $\Gamma_{+}(u^i,\gl,s)=e^{\pm\widehat{\phi}_{A_1}(u^i,\gl,s)_-}e^{\pm\widehat{\phi}_{A_1}(u^i,\gl,s)_+}$ and the coefficients are given by the following formulas:
\ben
\log \overline{c}_a=
\frac{1}{2}\int_{-{\bf 1}}^{t-\gl\,{\bf 1}} \W_{a,a} - 
\Omega(\phi_\ga(t,\gl,s)_+,\phi_\gb(t,\gl,s)_-) + 
\frac{1}{2}V\phi_\gb(t,\gl,s)_-^2
\een 
and 
\ben
\log \overline{c}_b=
\frac{1}{2}\int_{-{\bf 1}}^{t-\gl\,{\bf 1}} \W_{b,b} +
\Omega(\phi_\ga(t,\gl,s)_+,\phi_\gb(t,\gl,s)_-) + 
\frac{1}{2}V\phi_\gb(t,\gl,s)_-^2.
\een 
Let us rewrite the first formula in a slightly different way. We put $a=\ga+\gb$ and so we have $\W_{a,a}=\W_{\ga,\ga}+2\W_{\ga,\gb}+\W_{\gb,\gb}.$ Also we add and subtract the integral:
\ben
\frac{1}{2}\int_{-1}^{u^i-\gl} I^{(0)}_{A_1}(\xi,0,s)\bullet I^{(0)}_{A_1}(\xi,0,s) = 
\frac{1}{2}\int_{-1}^{u^i-\gl}\Big(\frac{1}{\sqrt{2(s-\xi)}}-\frac{1}{\sqrt{2(-\xi)}}  \Big)^2 d\xi.
\een

It should be clear that $\log \overline{c}_a$ is a sum of the following three functions. The first one:
\beq\label{ca:1}
\frac{1}{2}\int_{-{\bf 1}}^{t-\gl{\bf 1}} \W_{\ga,\ga} + 
\frac{1}{2}\int_{-1}^{u^i-\gl} I^{(0)}_{A_1}(\xi,0,s)\bullet I^{(0)}_{A_1}(\xi,0,s),
\eeq
the second one:
\beq\label{ca:2}
\int_{-{\bf 1}}^{t-\gl{\bf 1}} \W_{\ga,\gb} - 
\Omega(\phi_\ga(t,\gl,s)_+,\phi_\gb(t,\gl,s)_-),
\eeq
and the third one:
\beq\label{ca:3}
\frac{1}{2}V\phi_\gb(t,\gl,s)_-^2 +\frac{1}{2}\int_{-{\bf 1}}^{t-\gl{\bf 1}} \W_{\gb,\gb}-\frac{1}{2}\int_{-1}^{u^i-\gl}\Big(\frac{1}{\sqrt{2(s-\xi)}}-\frac{1}{\sqrt{2(-\xi)}}\Big)^2d\xi.  
\eeq
Notice that both \eqref{ca:2} and \eqref{ca:3} are single-valued near $\gl=u^i$. Indeed, let $\gamma$ be a small loop -- based at $(t,\gl)$ -- going around $(t,u^i)$, then the analytical continuation around $\gamma$ transforms $\Omega(\phi_\ga(t,\gl,s)_+,\phi_\gb(t,\gl,s)_-)$ into $\Omega(\phi_\ga(t,\gl,s)_+,\phi_{-\gb}(t,\gl,s)_-).$ However, the differential of $\Omega(\phi_\ga(t,\xi,s)_+,\phi_\gb(t,\xi,s)_-)$ is $\W_{\ga,\gb}(t-\xi\,{\bf 1})$ (see \leref{primitive}), so using the Stoke's theorem we get:
\beq\label{anal_gain}
\Omega(\phi_\ga(t,\gl,s)_+,\phi_\gb(t,\gl,s)_-) - \Omega(\phi_\ga(t,\gl,s)_+,\phi_{-\gb}(t,\gl,s)_-) = \int_{\gamma}\W_{\ga,\gb}.
\eeq 
The analytical continuation along $\gamma$, changes the value of the integral in \eqref{ca:2} also by $ \int_{\gamma}\W_{\ga,\gb}$, which implies that the function \eqref{ca:2} is single valued near $\gl=u^i$. Moreover, using \leref{primitive} and the Leibniz rule, we get that \eqref{ca:2} is independent of $t$ and $\gl$, so \eqref{ca:2} is just a constant. 

A similar argument proves that \eqref{ca:3} is also a constant. It remains to analyze \eqref{ca:1}. The first integral is an analytic function in $\gl$, because the cycle $\ga$ is invariant under the local monodromy. The exponent of the second integral is exactly the coefficient, which our general theory would prescribe for the $\W_{A_1}$-constraints of the ancestor potential of $A_1$-singularity (see Subsection \ref{step1}).  

For $\log \overline{c}_b$, the only difference is that in \eqref{ca:2} we have to replace $\gb$ with $-\gb$. Therefore, the regularity of \eqref{ui:constraints}
would follow from the $\W_{A_1}$ constraints for the ancestor potential of $A_1$-singularity, if we manage to prove that $\overline{c}_a= \overline{c}_b.$ 

The difference $\log \overline{c}_a - \log \overline{c}_b$ is equal to:
\beq\label{log:c}
\frac{1}{2}\int_{-{\bf 1}}^{t-\gl{\bf 1}} \W_{a,a} -\frac{1}{2}\int_{-{\bf 1}}^{t-\gl{\bf 1}} \W_{b,b} -2\Omega(\phi_\ga(t,\gl,s)_+,\phi_\gb(t,\gl,s)_-). 
\eeq
By definition $\gb=(a-b)/2$, so we have $-2\phi_\gb(t,\gl,s)_-=\phi_b(t,\gl,s)_- - \phi_a(t,\gl,s)_- .$ Therefore, using \leref{primitive} and the Stoke's theorem we get that the last term in \eqref{log:c} -- with the coefficient $(-2)$ included -- equals $\int_{\gamma_i} \W_{\ga,a},$ where $\gamma_i$ is a simple loop around the discriminant going around $(t,u^i)$. Using that $a=\ga+\gb$ we get
\ben
\int_{\gamma_i} \W_{\ga,a} = \int_{\gamma_i} \W_{\ga,\gb} = \frac{1}{2}\int_{\gamma_i}\W_{a,a},
\een
where for the last equality we used that $\int_{\gamma_i} \W_{\ga,\ga}=\int_{\gamma_i} \W_{\gb,\gb} = 0$. The first integral vanishes, because $\W_{\ga,\ga}$ is analytic near $\gl=u^i$, while the second one is zero thanks to \leref{period:beta}.  

Recall that the integration paths of the integrals in \eqref{log:c} have the form $C$ and $C\circ \gamma_0$, where $C$ is a path from $-{\bf 1}$ to $t-\gl{\bf 1}$ and $\gamma_0$ is a loop, based at $-{\bf 1}$, such that the parallel transport of $a$ along $\gamma_0$ is $b$. Therefore, \eqref{log:c} can be interpreted as: $\frac{1}{2}\oint_\gamma \W_{a,a}$, where $\gamma$ is the composition of the paths: $\gamma_0^{-1}\circ C^{-1}\circ\gamma_i\circ C.$ On the other hand, the cycle $a$ is invariant along $\gamma$. Recalling \leref{period:alpha} we get $\int_\gamma\W_{a,a}=0$. 

It remains only to prove $\W_{A_1}$ constraints for the ancestor potential of $A_1$ singularity. However, they follow from Subsection \ref{step1} and the $\W_{A_1}$ constraints for $\D_{\rm pt}$ -- see \coref{W:A1}. \thref{t1} is proved.  
\qed

\end{document}